\documentclass[letterpaper, 10 pt, journal, twoside]{ieeetran}
\usepackage{graphicx}
\usepackage{epstopdf}

\usepackage{cite}

\ifCLASSINFOpdf
\else
\fi
\usepackage[font={small}]{caption}
\usepackage{amsmath,amssymb}
\usepackage{amsfonts}
\usepackage{algorithm}
\usepackage{algorithmic}
\usepackage{array}
\usepackage{mathrsfs}
\usepackage{amsmath,amssymb,amsthm}
\usepackage{color}
\usepackage{multirow}
\usepackage{textcomp}
\usepackage[table]{xcolor}
\usepackage{amssymb}
\usepackage{tabularx}
\usepackage{booktabs}
\usepackage{cite}
\usepackage{graphicx}
\usepackage{caption}
\usepackage{subcaption} 
\usepackage{epstopdf}
\usepackage{upgreek}
\usepackage{comment}
\usepackage{hyperref}
\usepackage{amsmath,amssymb}
\usepackage{url}
\usepackage{cleveref}
\usepackage{booktabs}
\usepackage{diagbox}
\usepackage{pifont}
\usepackage{xcolor}

\newtheorem{theorem}{Theorem}

\newtheorem{lemma}{Lemma}
\newtheorem{corollary}{Corollary}
\newtheorem{remark}{Remark}
\newtheorem{proposition}{Proposition}

\newtheorem{problem}{Problem}

\definecolor{ResultSuccess}{HTML}{187F78}
\definecolor{ResultFailure}{HTML}{C94F54}

\newcommand{\ResultPass}{%
    \textcolor{ResultSuccess}{\ding{51}}}
\newcommand{\ResultFail}[1]{%
    \textcolor{ResultFailure}{\ding{55}}%
    \textsuperscript{\normalfont #1}}
    
\crefname{assumption}{assumption}{Assumptions}

\begin{document}

\title{Barrier-State-Based Safe Navigation in Complex Environments Using Look-Ahead Safety Corridors
}

\author{Yunlin Wei, Fangfei Li, \IEEEmembership{Member, IEEE}, and Junhui Zhang, \IEEEmembership{Member, IEEE}
\thanks{This work was supported by the National Natural Science Foundation of China under Grant 62603330. \emph{(Corresponding author:Junhui Zhang)}}
\thanks{Yunlin Wei and Junhui Zhang are with the School of Robotics and Automation at Nanjing University, Suzhou, Jiangsu 215163, China (e-mail: junhuiz@nju.edu.cn)}
\thanks{Fangfei Li is with the School of Mathematics, East China University of Science and Technology, Shanghai 200237, China (e-mail: lifangfei@ecust.edu.cn).}
\thanks{[Code] \protect\url{https://github.com/TRL-NJU/planning-guided-safety-corridor.git}}
\thanks{[Video] \protect\url{https://youtu.be/zIcdzXSXya4}}
 }

\maketitle
\thispagestyle{empty} 
\begin{abstract}
Safe navigation in cluttered environments requires a safety representation that captures the available motion space along the planned path. A single local safe region may restrict progress through narrow passages and continuous turns. This paper presents a planning-guided look-ahead safety corridor framework for barrier-state-based navigation. Local safe convex polytopes are constructed around collision-free seeds along the forward reference path. Their facet constraints are combined using softmin, and the local functions are fused using softmax with entropy-based adaptive compensation. The resulting safe set remains within the union of the local polytopes while reducing the shrinkage caused by fixed compensation. A Poisson boundary-value problem then defines a differentiable interior safety field while preserving the prescribed corridor boundary. This field is embedded into the robot dynamics through one geometric barrier state. We establish sufficient conditions for corridor invariance based on barrier-state boundedness and active-corridor validity. Navigation simulations cover cluttered layouts, narrow passages, and moving obstacles. A differential-drive ground vehicle further demonstrates navigation through narrow passages and among obstacles absent from the prior map.

\end{abstract}


\section{Introduction} 

\IEEEPARstart{S}{afe}  navigation in cluttered environments requires a robot to plan feasible paths and maintain a safety margin while moving along them. This requirement becomes particularly demanding in narrow passages and around successive turns, where feasible motion strongly depends on the geometry ahead. Recent planning methods have improved local trajectory generation, perception-aware replanning, and navigation in unknown environments~\cite{zhou2021ego,zhou2021raptor,tordesillas2022faster}. However, a collision-free reference path only specifies a route toward the goal and does not explicitly describe the surrounding space available to feedback control. If the safety representation is limited to the immediate area around the robot, the robot might be overly constrained before the robot is about to turn or enter a passage. Therefore, connecting the planned path with a geometric representation for safe feedback is essential for sustained navigation in constrained environments.

%CBF
Control barrier functions (CBFs) and their higher-order extensions have been used to enforce state constraints through forward invariance to guarantee system safety~\cite{ames2016control,xiao2021high}. Their applications extend beyond navigation. For example, Zhang et al. combine sparse feedback with CBF constraints for frequency regulation in inverter-intensive microgrids~\cite{zhang2023novel}. For stochastic systems with data-driven state estimators, conformal prediction has been incorporated into sampled-data risk-aware CBF design to account for estimation uncertainty~\cite{zhang2025conformal}. For navigation, geometry-dependent CBFs have been constructed using differentiable optimization~\cite{dai2023safe} and quasi-conformal mappings~\cite{notomista2025reactive}. Velocity obstacle-based CBFs address dynamic collision avoidance in velocity space~\cite{huang2025dynamic}. However, CBF-based quadratic programs can admit undesired equilibria, so safety alone does not generally ensure progress toward the goal~\cite{tan2024equilibria}. The Barrier State (BaS) framework offers a complementary way to incorporate safety into control design~\cite{almubarak2025barrier}. It augments the system with the dynamics of a transformed safety function and relates safety to the boundedness of the resulting barrier state~\cite{almubarak2025barrier}. Discrete barrier states have also been incorporated into differential dynamic programming~\cite{almubarak2022ddp}. For navigation, the underlying safety function must still be constructed from the reference path and obstacle geometry. Its admissible set determines the available motion space, while its values and derivatives enter the augmented dynamics. Constructing this differentiable safety function is thus a key step in connecting path planning with BaS-based control.

Safe corridors represent the available motion space through collections of obstacle-free regions. Iterative Regional Inflation by Semidefinite Programming (IRIS) was introduced to compute large convex regions through alternating separation and ellipsoid inflation~\cite{deits2015iris}. Overlapping convex polyhedra have been used to form safe flight corridors for trajectory generation~\cite{liu2017corridors}. Fast Iterative Region Inflation (FIRI) was developed to improve construction efficiency while retaining prescribed seed sets~\cite{wang2025firi}. Convex regions also support geometrically constrained trajectory optimization~\cite{wang2022gcopter} and motion planning through graphs of convex sets~\cite{marcucci2023motion}. More recently, PathCover was proposed to construct overlapping convex polytopes directly from finite point clouds using Randomized Iterative Space Partitioning. It establishes path coverage with respect to the supplied point set~\cite{narkhede2026pathcover}. Corridor-based CBF and CLF constraints have also been incorporated into model predictive contouring control~\cite{mohammad2025corridor}. In another approach, CBF conditions were converted into control barrier corridors that characterize safe goal regions~\cite{arslan2026corridors}. Through such methods, geometric safety is combined with input constraints and reference trajectory selection.

%smooth
Representing a corridor composed of local safe convex polytopes (SCPs) through a single safety function requires both half-space intersections and region unions. These set operations can be represented exactly using minimum and maximum operators, but the resulting functions are generally nonsmooth. Smooth approximations have also been used to construct closed-form CBF candidates for polytope navigation~\cite{molnar2025polytopes}. Soft-minimum and soft-maximum barrier functions have been developed for safety under actuation constraints~\cite{rabiee2025soft}. Time-varying soft-maximum functions have also been used to combine recently perceived local safe sets in unmapped and dynamic environments~\cite{safari2025time}. However, the conservatism introduced by smoothing can reduce the space needed to traverse narrow passages. Fixed softmax compensation corrects for the largest possible aggregation offset, introducing unnecessary shrinkage even where one local function dominates. Meanwhile, the spatial variation of the fused function remains influenced by the individual SCP functions and their overlap. Reducing compensation alone does not remove this dependence on the local representation.

%poisson
Corridor geometry and interior field design are distinct issues. In the BaS framework, the interior values and gradients matter in addition to the safety boundary. Direct fusion leaves these quantities dependent on the local functions and their overlap. Poisson boundary-value problems offer a way to construct a field
on a prescribed domain. Bahati et al. construct Poisson safety functions from perception data~\cite{bahati2025dynamic}. Subsequent work incorporates robot geometry and motion domains into predictive safety filtering~\cite{bena2025geometry}. Risk-aware safety filters further distinguish safety functions from guidance fields~\cite{bahati2026risk}. These studies motivate reconstruction of an interior field after corridor acceptance.
The field is then specified by the accepted boundary and a
prescribed source, rather than by the explicit algebraic fusion
of individual facet functions.

%contribution
We propose a  planning-guided construction of look-ahead safety
corridors and their interior fields. Local SCP functions are fused using softmax with entropy-based adaptive compensation. Then a Poisson field is reconstructed within the accepted corridor and embedded through a single barrier state. The safety result is conditional on barrier-state boundedness and active-corridor validity.

The main contributions are as follows:
\begin{enumerate}
    \item We develop a differentiable look-ahead corridor formulation with adaptive compensation. Its safe set contains the fixed-compensation approximation and remains within the union of the local safe regions, reducing unnecessary shrinkage without compromising geometric containment.
    \item We construct a Poisson safety field over the accepted
    corridor. This separates corridor geometry from interior field design. The field preserves the prescribed boundary and provides the values and spatial derivatives required for BaS embedding.
    \item We embed the field using one geometric barrier state, independently of the number of local polytopes and facets. We establish conditional corridor invariance and evaluate the construction in static and dynamic navigation simulations. Ground-vehicle experiments demonstrate deployment in narrow passages and among previously unknown obstacles.
\end{enumerate}

\begin{figure*}[!t]
\vspace{5mm}
    \centering 
    \includegraphics[width=\textwidth]{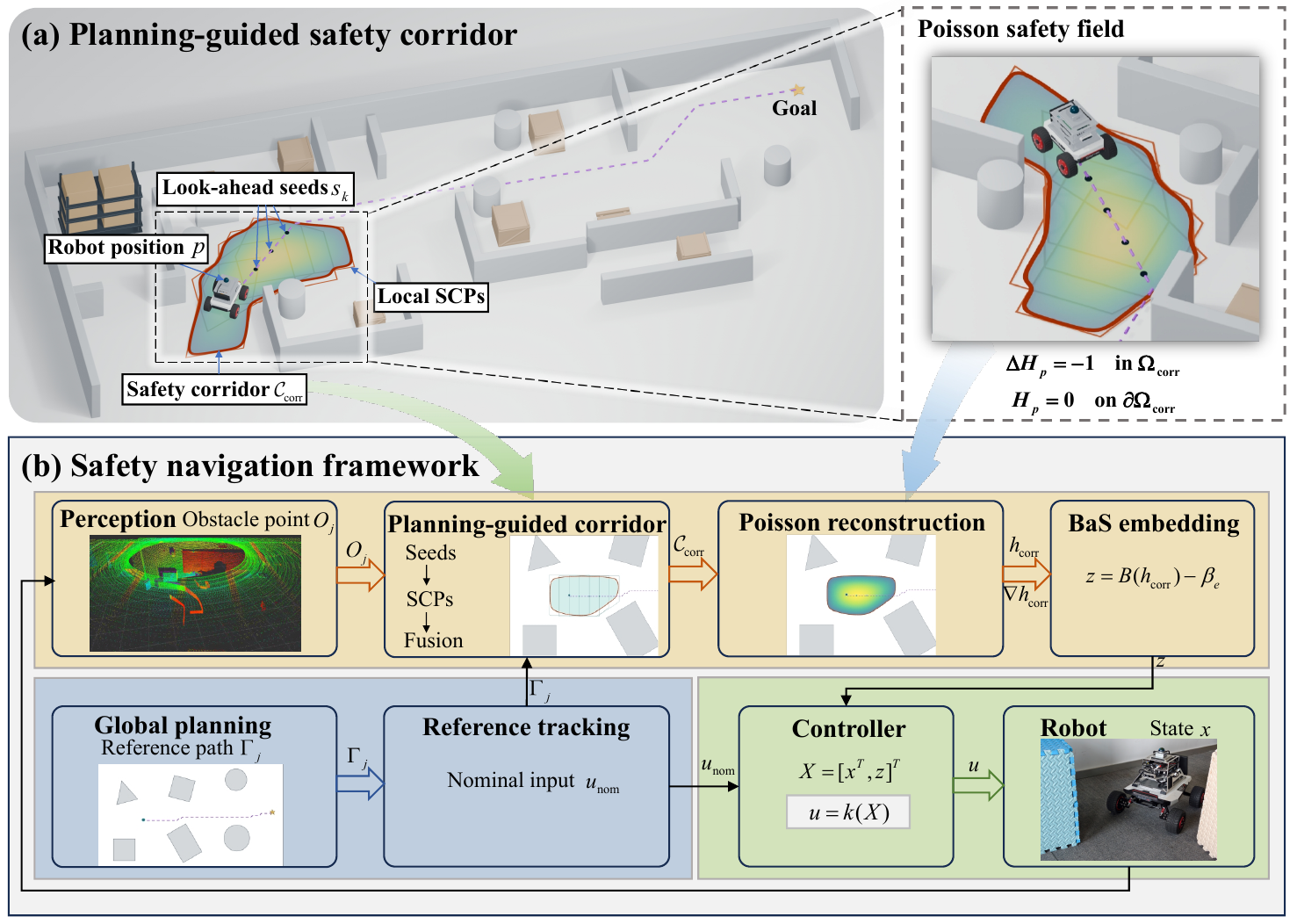}
    \caption{Closed-loop architecture of the proposed planning-guided Barrier-State navigation framework.}
\label{fig:framework}
\end{figure*}

\section{Preliminaries}
Throughout this paper, $\|\cdot\|$ denotes the Euclidean norm. For a set $\mathcal{C}$, $\operatorname{Int}(\mathcal{C})$ and $\partial\mathcal{C}$ denote its interior and boundary, respectively.
\subsection{Safe Set and Barrier Function}
\label{subsec:cbf}
Consider a robot described by the nonlinear control-affine system:
\begin{equation}
    \dot{x}=f(x)+g(x)u,\qquad x(0)=x_0,
    \label{eq:control_affine_system}
\end{equation}
where $x\in\mathbb{R}^{n}$ is the system state and $u\in\mathcal{U}\subseteq\mathbb{R}^{m}$ is the control input. The $f:\mathbb{R}^n \to \mathbb{R}^n$, $g:\mathbb{R}^n \to \mathbb{R}^{n \times m}$ are assumed to be locally Lipschitz continuous. Following the general BaS setting~\cite{almubarak2025barrier}, $x$ contains the variables required by the selected model, such as position, orientation, velocity, or actuator states. The workspace position is extracted by
\begin{equation}
    p=\pi(x)\in\mathbb R^d,
    \label{eq:position_mapping}
\end{equation}
where $\pi$ is continuously differentiable and $d$ is the workspace dimension. Here, $d=2$ for planar navigation and $d=3$ for spatial navigation. The geometric construction uses $p$, while the dynamics of $x$ are used for the state-space embedding.

Let $h:\mathbb{R}^n \to \mathbb{R}$ be  continuously differentiable and define
\begin{equation}
\label{eq: safe set}
    \mathcal{C}=\{x \in \mathbb{R}^n: \ h(x) \geq 0\}.
\end{equation}
We consider a defining function $h$ for which the interior and boundary are respectively given by $\operatorname{Int}(\mathcal C)=\{x\mid h(x)>0\}$ and $\partial\mathcal C=\{x\mid h(x)=0\}$. Safety means that a closed-loop trajectory initialized in $\mathcal C$ remains in $\mathcal C$. The BaS construction below uses initial states in $\operatorname{Int}(\mathcal C)$.

The Lie derivatives of $h$ are
\begin{equation}
    L_fh(x)=\nabla h(x)^\top f(x),\qquad
    L_gh(x)=\nabla h(x)^\top g(x).
    \label{eq:lie_derivatives}
\end{equation}
Thus, along~\eqref{eq:control_affine_system}, $\dot{h}=L_fh+L_gh\,u$.

Let $B:(0,\infty)\rightarrow\mathbb{R}^+$ be a continuously differentiable and strictly decreasing function satisfying
\begin{equation}
\label{eq: barrier_singular}
    \lim_{h\rightarrow 0^+}B(h)=+\infty .
\end{equation}
Typical choices are $B(h)=1 / h$ and $B(h)=-\ln(h / (1+h))$.

Define the transformed safety variable $\beta(x):=B(h(x))$. As the safety margin approaches zero, $\beta(x)$ diverges. A finite upper bound on $\beta(x(t))$ therefore excludes approach to the zero level set of $h$. This property is the basis of the Barrier-State formulation~\cite{almubarak2025barrier}.

\subsection{Barrier States}

The Barrier-State (BaS) approach incorporates the dynamics of a transformed safety variable into the original system~\cite{almubarak2025barrier}. Let $\beta_e$ be a finite fixed reference value. Under consistent initialization, the Barrier State satisfies
\begin{equation}
    z(t)=B(h(x(t)))-\beta_e.
    \label{eq:bas_consistency}
\end{equation}
When an admissible equilibrium $x_e$ lies in the safe interior, choosing $\beta_e=B(h(x_e))$ gives $z=0$ at $x=x_e$.

With the augmented state $X=[x^\top,z]^\top$, the safety-embedded system is
\begin{equation}
    \dot X
    =
    \underbrace{
    \begin{bmatrix}
        f(x)\\
        f_b(x,z)
    \end{bmatrix}}_{F(X)}
    +
    \underbrace{
    \begin{bmatrix}
        g(x)\\
        g_b(x,z)
    \end{bmatrix}}_{G(X)}u,
    \label{eq:augmented_system}
\end{equation}
where $f_b$ and $g_b$ define the Barrier-State dynamics. Their expressions and the initialization that preserves
\eqref{eq:bas_consistency} are given in Appendix~\ref{app:bas_dynamics}.

\subsection{Obstacle Representation}
\label{subsec:obstacle}
The robot is represented geometrically by its reference position $p$ defined in~\eqref{eq:position_mapping}. A clearance parameter $\rho>0$ accounts for a conservative bound on its physical extent, the prescribed safety margin, and bounded errors in the obstacle representation. This converts geometric collision checking into a point-clearance condition without tying the construction to a particular dynamic model.

Let
\begin{equation}
    \mathcal O  = \{ o_{1},o_{2},\ldots,o_{M} \} \subset\mathbb{R}^{d}
    \label{eq:perceived_obstacle_set}
\end{equation}
be the latest set of perceived obstacle points. For a candidate position $q\in\mathbb{R}^{d}$, define
\begin{equation}
    \operatorname{dist}(q,\mathcal{O})
    :=\min_{o\in\mathcal{O}}\|q-o\|.
    \label{eq:obstacle_distance}
\end{equation}
When $\mathcal{O}=\varnothing$, this distance is defined as $+\infty$. A strictly positive clearance margin corresponds to
$\operatorname{dist}(q,\mathcal{O})>\rho$.

The obstacle representation is assumed to be conservative at the selected clearance scale. In particular, finite samples must resolve the relevant obstacle geometry, with bounded sampling and perception errors included in $\rho$. Separation from sampled points alone does not establish clearance from unrepresented obstacle geometry. The set $\mathcal{O}$ is replaced when new sensing data become available. 

\subsection{Softmin and Softmax}
\label{subsec:softmin_softmax}
Smooth approximations of minimum and maximum operators allow several scalar constraints to be combined into one differentiable function. For $\kappa>0$ and scalar arguments $h_1,\ldots,h_N$, the log-sum-exp construction gives~\cite{rabiee2025soft}: %log-sum-exp
\begin{equation}
    \begin{aligned}
        \operatorname{softmin}_{\kappa}(h_1,\ldots,h_N)
        &=
        -\frac{1}{\kappa}
        \ln\left(\sum_{i=1}^{N}e^{-\kappa h_i}\right),\\
        \operatorname{softmax}_{\kappa}(h_1,\ldots,h_N)
        &=
        \frac{1}{\kappa}
        \ln\left(\sum_{i=1}^{N}e^{\kappa h_i}\right)
        -\frac{\ln N}{\kappa}.
    \end{aligned}
    \label{eq:softmin_softmax}
\end{equation}

These functions satisfy
\begin{equation}
    \begin{aligned}
        \min_i h_i-\frac{\ln N}{\kappa}
        &\leq
        \operatorname{softmin}_{\kappa}(h_1,\ldots,h_N)
        \leq\min_i h_i,\\
        \max_i h_i-\frac{\ln N}{\kappa}
        &\leq
        \operatorname{softmax}_{\kappa}(h_1,\ldots,h_N)
        \leq\max_i h_i.
    \end{aligned}
    \label{eq:smooth_extrema_bounds}
\end{equation}
Thus, both approximations approach their corresponding extrema as $\kappa\rightarrow\infty$. A larger $\kappa$ reduces the approximation error but produces sharper transitions near competing arguments.  If the arguments are continuously differentiable functions of position, their compositions in~\eqref{eq:softmin_softmax} are also continuously differentiable.

\begin{proposition}[Conservative set composition]
\label{prop1}
Let $h_i:\mathbb{R}^{d}\rightarrow\mathbb{R}$ be continuous
and define
$\mathcal{C}_i:=\{p\in\mathbb{R}^{d}\mid h_i(p)\geq0\}$.
Then
% \begin{equation}
%     \begin{aligned}
%         \operatorname{softmin}_{\kappa}
%         (h_1(p),\ldots,h_N(p))\geq0
%         &\ \Longrightarrow\
%         p\in\bigcap_{i=1}^{N}\mathcal{C}_i,\\
%         \operatorname{softmax}_{\kappa}
%         (h_1(p),\ldots,h_N(p))\geq0
%         &\ \Longrightarrow\
%         p\in\bigcup_{i=1}^{N}\mathcal{C}_i.
%     \end{aligned}
%     \label{eq:smooth_set_containment}
% \end{equation}
\begin{equation}
    \begin{aligned}
        \{p\mid \operatorname{softmin}_{\kappa}
        (h_1(p),\ldots,h_N(p))\geq0\}
        &\subseteq \bigcap_{i=1}^{N}\mathcal C_i,\\
        \{p\mid \operatorname{softmax}_{\kappa}
        (h_1(p),\ldots,h_N(p))\geq0\}
        &\subseteq \bigcup_{i=1}^{N}\mathcal C_i.
    \end{aligned}
\end{equation}

\end{proposition}

\begin{proof}
The first implication follows from
$\operatorname{softmin}_{\kappa}\leq\min_i h_i$,
and the second follows from
$\operatorname{softmax}_{\kappa}\leq\max_i h_i$.
\end{proof}

Softmin therefore provides an inner approximation of an intersection, while the compensated softmax provides an inner approximation of a union. The fixed compensation in~\eqref{eq:softmin_softmax} serves as a reference construction. Section IV-A introduces the adaptive compensation used for corridor fusion and establishes its set-containment property separately.

\section{Problem Formulation}
\label{sec:problem}
Consider the robot in~\eqref{eq:control_affine_system} navigating a workspace $\mathcal W\subset\mathbb R^d$ toward a prescribed goal. Let $\{t_j\}_{j=0}^{\infty}$ denote the times at which accepted corridors are activated, with $t_0=0$, $t_{j+1}>t_j$, and $t_j\to\infty$. The index $j$ identifies the active corridor. At $t_j$, let $p_j=\pi(x(t_j))$ denote the robot position, $\mathcal O_j$ the latest available obstacle point set, and $\Gamma_j$ the available reference path. The accepted corridor and its Poisson field are held fixed on $[t_j,t_{j+1})$. These activation times need not coincide with every sensor measurement or control step. Using the clearance parameter defined in Section~\ref{subsec:obstacle}, the perceived collision-free space is
\begin{equation*}
    \mathcal F_j :=\left\{q\in\mathcal W\;\middle|\;
    \operatorname{dist}(q,\mathcal O_j)\geq\rho\right\}.
\end{equation*}
This set describes the geometry available at $t_j$. Holding this representation fixed between updates does not imply that the physical environment remains unchanged.

\noindent\textbf{Planning-Guided Corridor Construction:} 
The reference path provides a route toward the goal, but its forward portion may become infeasible under the latest obstacle information. The geometric task is to identify a feasible forward portion and construct a bounded local corridor around it. Let $\Gamma_j^{e}$ denote the selected forward reference segment, including a collision-free connection from $p_j$ to the reference path when needed.

The corridor is represented by a continuously differentiable function $H_{\mathrm{corr},j}:\mathbb{R}^{d}\rightarrow\mathbb{R}$,
with
\begin{equation*}
    \mathcal C_{\mathrm{corr},j}:=\left\{q\in\mathbb R^d\;\middle|\;
    H_{\mathrm{corr},j}(q)\geq0\right\}
\end{equation*}
A valid corridor is required to be connected and to satisfy
\begin{equation}
    \begin{aligned}
        p_j
        &\in\operatorname{Int}
        (\mathcal{C}_{\mathrm{corr},j}),\\
        \Gamma_j^{e}
        &\subseteq\operatorname{Int}
        (\mathcal{C}_{\mathrm{corr},j}),\\
        \mathcal{C}_{\mathrm{corr},j}
        &\subseteq\mathcal{F}_j.
    \end{aligned}
    \label{eq:corridor_requirements}
\end{equation}

These conditions require the corridor to contain the current position and the retained reference segment without extending beyond the perceived free space. They are acceptance requirements for the final corridor, not consequences of seed-point safety alone.

\noindent\textbf{Safety Field Construction:}
The zero level set of $H_{\mathrm{corr},j}$ defines the corridor boundary, whereas its interior values depend on the local SCPs and their fusion. The field-construction task is to obtain a Poisson safety field $H_{P,j}$ on the accepted corridor, with positive values in its interior and zero values on its boundary. The construction must preserve the corridor geometry and provide the spatial derivatives required for the BaS embedding. Section IV-B specifies the boundary-value problem and its regularity conditions.

\noindent\textbf{Safe Navigation:}
During $[t_j,t_{j+1})$, the active corridor and its Poisson field are held fixed. The control objective is to follow the retained
forward reference toward the goal while keeping the robot inside the active corridor. The Poisson field is mapped to the state space through $p=\pi(x)$ and incorporated into the safety-embedded system of Section II-B. The control input is not restricted to a particular controller structure; the required closed-loop safety property is established through barrier-state boundedness in Section V.

\begin{problem}[Planning-Guided Safe Navigation]
Given the current state $x(t_j)$, obstacle information $\mathcal O_j$, and reference path $\Gamma_j$, construct a forward reference segment $\widetilde\Gamma_j$, a corridor function $H_{\mathrm{corr},j}$ satisfying~\eqref{eq:corridor_requirements}, and the corresponding Poisson safety field $H_{P,j}$.

Embed the field into the BaS framework and characterize the conditions under which an admissible feedback $u(t)\in\mathcal U$ keeps the robot inside the active corridor while tracking the retained reference segment.
\end{problem}

% \begin{problem}[Planning-Guided Safe Navigation]
% Given the current state $x(t_j)$, obstacle information $\mathcal{O}_j$, and reference path $\Gamma_j$, construct a forward reference segment $\Gamma_j^{e}$, a corridor function $H_{\mathrm{corr},j}$ satisfying~\eqref{eq:corridor_requirements}, and the corresponding Poisson safety field $H_{P,j}$. Using this field, determine an admissible closed-loop input that supports progress toward the goal while satisfying
% \begin{equation*}
%     u(t)\in\mathcal U, \qquad
%     \pi(x(t))\in\operatorname{Int}(\mathcal C_{\mathrm{corr},j}), \quad 
%     t\in[t_j,t_{j+1}).
% \end{equation*}
% The construction is repeated using updated perception and reference-path information. If no candidate satisfies the corridor acceptance requirements, construction failure is reported. The safety analysis assumes that each active corridor remains collision-free over the interval in which it is used.
% \end{problem}

\section{Planning-Guided Look-Ahead Safety Corridor Construction}
\label{sec:corridor_construction}
This section derives a workspace safety representation from the reference path and the latest obstacle information. Section IV-A first combines local Safe Convex Polytopes (SCPs) along the reference path to obtain a continuously differentiable corridor function. Section IV-B then constructs a Poisson field inside the corridor. For clarity, the update index $j$ is omitted in this section because the construction is described for a single update.

\subsection{Differentiable Safety Corridor Construction}
\label{subsec:differentiable_corridor}
The corridor construction contains four steps. First, collision-free seed points are selected from the forward portion of the reference path. Second, an SCP is constructed around each seed point. Third, the facet functions of each SCP are combined into one differentiable local function. Finally, the accepted local functions are fused into a single corridor function using adaptive compensation. The overall process is illustrated in Figs.~\ref{fig:scp_construction} -~\ref{fig:whole corridor} .
\subsubsection{Look-Ahead Seed Selection}
Let
\begin{equation*}
    \Gamma=\{\gamma_1,\gamma_2,\ldots,\gamma_L\}
    \subset\mathbb{R}^{d}
\end{equation*}
denote the ordered reference path, where $\gamma_\ell$ is the $\ell$-th waypoint. The construction accepts an ordered path from a global planner. In the experiments, we use Jump Point Search (JPS), a grid-based path-planning method~\cite{harabor2011online}.
At each corridor-construction attempt, the waypoint closest to the current robot position is identified. Candidate seed points are then sampled in order from the forward part of the path. Each candidate is checked using the latest obstacle information, and the sampling process stops when the next candidate is unsafe or the prescribed look-ahead range is reached.
The waypoint closest to the current position $p=\pi(x)$ is determined by

\begin{equation*}
    \ell_0 = \underset{\ell\in\{1,\ldots,L\}}{\arg\min}\|p-\gamma_\ell\|,
\end{equation*} 
Starting from $\gamma_{\ell_0}$, candidate points are sampled forward along $\Gamma$ with a spatial interval $\Delta s$. The current position is included as the first seed point,
\begin{equation*}
    s_1=p.
\end{equation*}
A subsequent candidate $s_k$ is accepted if 
\begin{equation}
    \operatorname{dist}(s_k,\mathcal O)
    :=
    \min_{o\in\mathcal O}\|s_k-o\|>\rho,
    \label{eq:dist_sk_o}
\end{equation}
where $\rho$ includes the robot footprint, the required safety margin, and the bounded geometric uncertainty. When $\mathcal O=\varnothing$, the distance is defined as $\operatorname{dist}(s_k,\varnothing)=+\infty$.

The candidates are examined sequentially in the forward direction of $\Gamma$. Sampling terminates at the first point that violates~\eqref{eq:dist_sk_o}, or when no further point is available within the look-ahead range. The resulting seed sequence is
\begin{equation*}
    \mathcal S=\{s_1,s_2,\ldots,s_N\}
    \subset\mathbb{R}^{d},
    \qquad s_1=p,
\end{equation*}
where $N$ is determined by the current path and obstacle geometry. Every seed in $\mathcal S$ is therefore collision-free with respect to the current obstacle set.

\subsubsection{Local SCP Construction and Differentiable Representation}
\label{subsubsec:local_scp}
For each collision-free seed point $s\in\mathbb{R}^{d}$, let $\mathcal D(s)\subset\mathbb{R}^{d}$ be a bounded convex polytope satisfying $s\in\operatorname{Int}(\mathcal D(s))$. Let $\mathcal F_{\mathcal D}$ denote the facet-function set associated with the boundary of $\mathcal D(s)$. Obstacle points outside $D(s)$ may still affect the clearance of positions inside it. We therefore select
\begin{equation}
    \mathcal O_s = \{o\in\mathcal O:
    \operatorname{dist}(o,D(s))\le\rho\},
    \label{eq:local_obstacle_set}
\end{equation}
where $\operatorname{dist}(o,D(s)) =\inf_{q\in D(s)}\|o-q\|$. Only the obstacle selection range is enlarged. The resulting SCP is still limited within $D(s)$. If $\mathcal O_s=\varnothing$, the local construction directly returns $\mathcal P(s)=\mathcal D(s)$ and $\mathcal F_s=\mathcal F_{\mathcal D}$. Otherwise, the points in $\mathcal O_s$ are separated iteratively from the seed point.

\begin{figure}[!tbp]
    \centering
    \begin{subfigure}[b]{\columnwidth}
        \centering
        \includegraphics[width=\textwidth]{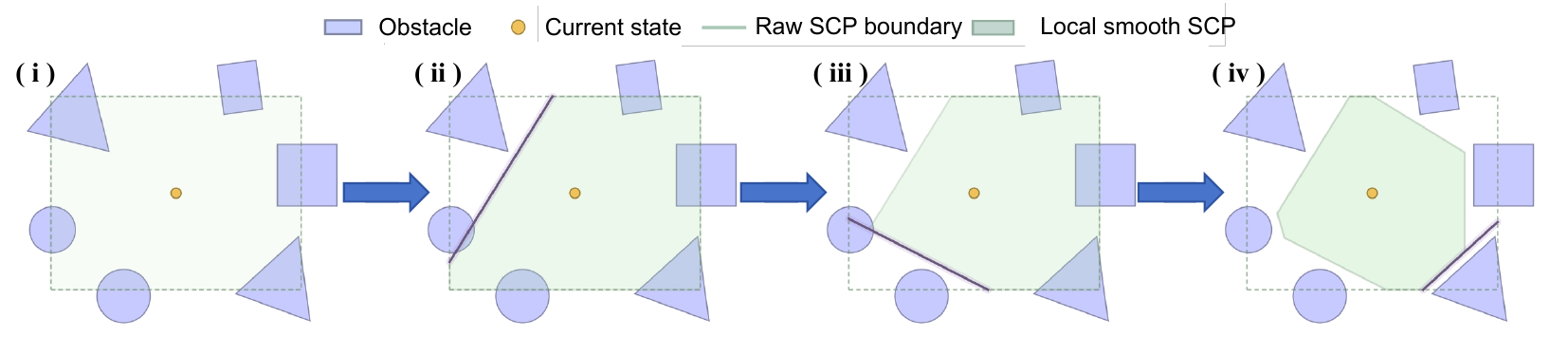}
        \caption{Construction of a 2D SCP among mixed-shape obstacles}
        \label{fig:2D_scp_construction}
    \end{subfigure}
    \begin{subfigure}[b]{\columnwidth}
        \centering
        \includegraphics[width=\textwidth]{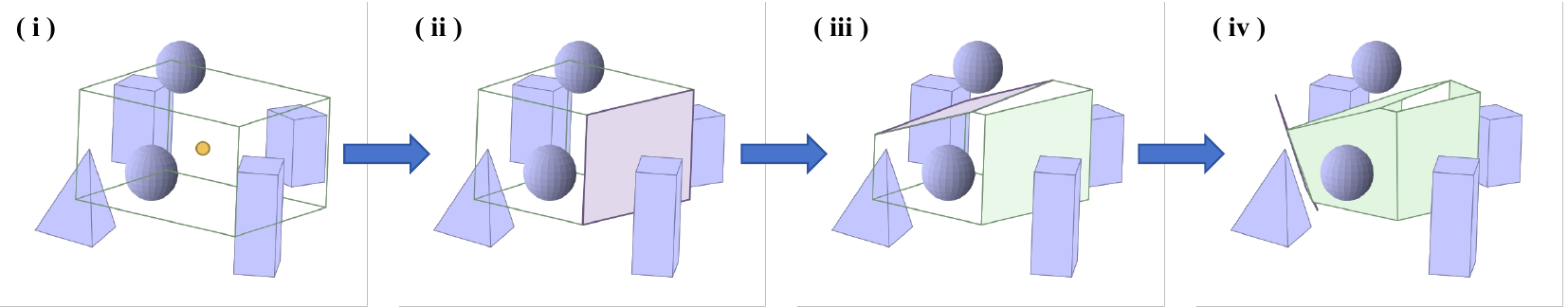}
        \caption{Construction of a 3D SCP in a cluttered environment}
        \label{fig:3D_scp_construction}
    \end{subfigure}
    \caption{Iterative construction of an SCP in two- and three-dimensional workspaces. Panels (i)--(iv) show the bounded construction domain, two intermediate separating hyperplanes, and the final SCP. At each iteration, the closest point in the remaining obstacle set determines the active separator, and the half-space containing the seed point is retained.}
    \label{fig:scp_construction}
\end{figure}

As shown in Fig.~\ref{fig:scp_construction}, the closest point in the remaining obstacle set is selected at each iteration. A separating hyperplane is constructed normal to the direction from the obstacle point to the seed. The half-space containing the seed is retained, while obstacle points already separated by the current hyperplane are removed from later iterations. The construction produces a convex polygon for $d=2$ and a convex polyhedron for $d=3$.

Initialize the candidate obstacle set as
$\mathcal C_0=\mathcal O_s$. At iteration $r$, let $\mathcal C_{r-1}\subseteq\mathcal O_s$ denote the remaining candidate obstacle set. The closest obstacle point, the corresponding unit normal, the retained half-space, and its facet function are defined as
\begin{equation}
\begin{aligned}
    o^{(r)} &= \underset{o\in\mathcal C_{r-1}}{\arg\min} \|s-o\|,
    \\[1mm]
    a_{r} &= \frac{s-o^{(r)}} {\|s-o^{(r)}\|},
    \\[1mm]
    \mathcal H_{r} &= \left\{p\in\mathbb{R}^{d} \middle|\, a_{r}^{\top}(p-o^{(r)})-\rho \geq 0\right\},
    \\[1mm]
    h_{r}(p) &= a_{r}^{\top} (p-o^{(r)})-\rho.
\end{aligned}
\label{eq:scp_geometric_step}
\end{equation}

Since $a_r$ points from $o^{(r)}$ toward $s$ and $\|s-o^{(r)}\|>\rho$, the seed point lies in the interior of $\mathcal H_r$. The candidate obstacle set is updated according to
\begin{equation}
    \mathcal C_{r} = \left\{o\in\mathcal C_{r-1} \middle|\, a_{r}^{\top}(o-o^{(r)})>0\right\}.
    \label{eq:scp_candidate_update}
\end{equation}

At least one obstacle point is removed at each iteration. The construction therefore terminates after a finite number of iterations.

After $R_s$ iterations, the SCP and its complete facet-function set are
\begin{equation}
\begin{aligned}
    \mathcal P(s)
    &=
    \mathcal D(s)\cap
    \bigcap_{r=1}^{R_s}\mathcal H_r,
    \\
    \mathcal F_s
    &=
    \mathcal F_{\mathcal D}
    \cup
    \{h_r(p)\}_{r=1}^{R_s}.
\end{aligned}
\label{eq:scp_definition}
\end{equation}

For any obstacle point $o$ removed at iteration $r$, the removal rule gives $a_r^\top(o-o^{(r)})\le0$. For every $q\in P(s)\subseteq H_r$,
\begin{equation*}
    a_r^\top(q-o) =
    a_r^\top(q-o^{(r)}) -a_r^\top(o-o^{(r)})
    \ge\rho.
\end{equation*}
Since $\|a_r\|=1$, this implies $\|q-o\|\ge\rho$. Every selected point is removed by some separator before termination. For an unselected point, $\operatorname{dist}(o,D(s))>\rho$. Since $P(s)\subseteq D(s)$, it also satisfies $\operatorname{dist}(o,P(s))>\rho$. Consequently,
\begin{equation*}
    \operatorname{dist}(q,\mathcal O)\ge\rho,
    \qquad q\in P(s).
\end{equation*}
This result applies to the supplied obstacle points. Physical clearance further relies on the conservative representation in Section~\ref{subsec:obstacle}.

Applying this construction to the $k$-th seed point $s_k$ gives
\begin{equation*}
    \mathcal P_k=\mathcal P(s_k),
    \qquad
    \mathcal F_k=
    \{h_{ki}(p)\}_{i=1}^{N_k},
    \qquad
    N_k=|\mathcal F_k|.
\end{equation*}

The facet functions of $\mathcal P_k$ are then combined using the softmin operator in~\eqref{eq:softmin_softmax}:
\begin{equation}
    H_k(p)=
    -\frac{1}{\kappa}
    \ln\left(
    \sum_{i=1}^{N_k}
    e^{-\kappa h_{ki}(p)}
    \right),
    \qquad \kappa>0.
    \label{eq:local_softmin1}
\end{equation}

The corresponding differentiable local region is
\begin{equation}
    \widetilde{\mathcal P}_k
    =
    \left\{
    p\in\mathbb{R}^{d}
    \,\middle|\,
    H_k(p)\geq0
    \right\}.
    \label{eq:smooth_local_safe_region}
\end{equation}

By Proposition~\ref{prop1},
\begin{equation*}
    H_k(p)\geq0
    \quad\Longrightarrow\quad
    h_{ki}(p)\geq0,
    \qquad i=1,\ldots,N_k.
\end{equation*}

Therefore,
\begin{equation}
    \widetilde{\mathcal P}_k
    \subseteq
    \mathcal P_k.
    \label{eq:local_smooth_containment}
\end{equation}

Hence, the multiple facet inequalities are represented by one continuously differentiable function without enlarging the original SCP.

The local regions are examined in the order of their seed points. For $k=2,\ldots,N$, the $k$-th region is retained only if
\begin{equation}
    \operatorname{Int}
    (\widetilde{\mathcal P}_{k-1})
    \cap
    \operatorname{Int}
    (\widetilde{\mathcal P}_k)
    \neq\varnothing.
    \label{eq:local_region_overlap}
\end{equation}

If this condition is not satisfied, expansion stops at the previous region. Let $N_c\leq N$ denote the number of retained regions. Their interiors form an overlapping sequence along the reference path. These regions are then fused into a single safety function.
Since fusion gives an inner approximation of their union, the resulting corridor is checked for reference coverage and connectivity before use.

\begin{figure}[!tbp]
    \centering
    \begin{subfigure}[b]{\columnwidth}
        \centering
        \includegraphics[width=\textwidth]{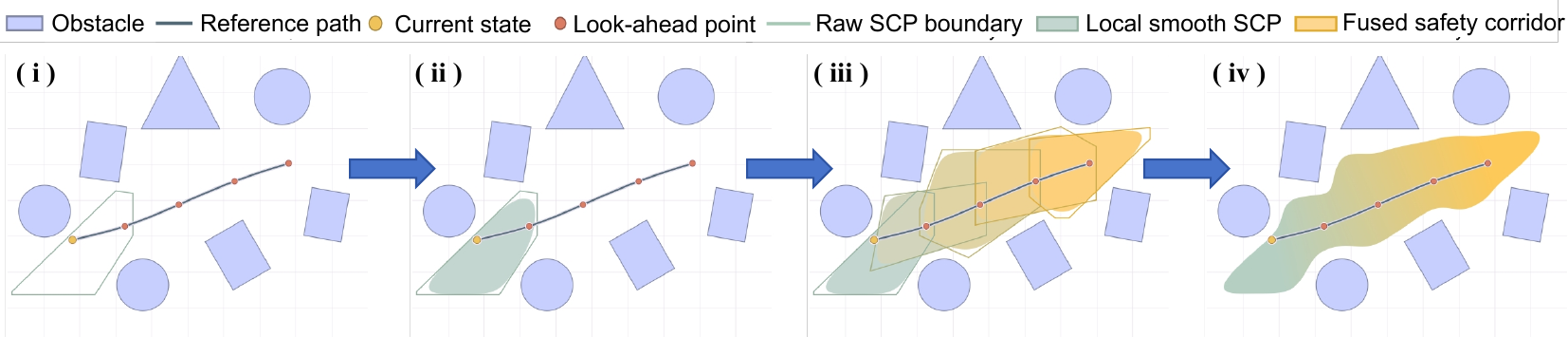}
        \caption{Construction of the planning-guided look-ahead safety corridor in two dimensions}
        \label{fig:2D_corridor_construction}
    \end{subfigure}
    \begin{subfigure}[b]{\columnwidth}
        \centering
        \includegraphics[width=\textwidth]{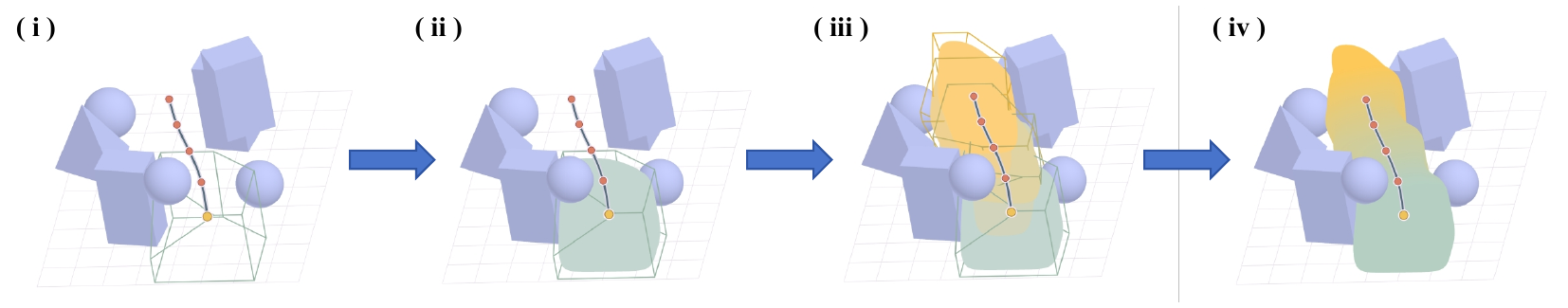}
        \caption{Construction of the planning-guided look-ahead safety corridor in three dimensions}
        \label{fig:3D_corridor_construction}
    \end{subfigure}
    \caption{Columns (i)--(iv) show: (i) construction of the SCP at the current position; (ii) differentiable representation of the local SCP; (iii) construction of local SCPs along the reference path; and (iv) fusion of the local functions into the corridor representation.}
    \label{fig:corridor_construction}
\end{figure}

Fig.~\ref{fig:corridor_construction} summarizes the progression from a local SCP to the fused look-ahead corridor.

\begin{figure*}[!t]
\vspace{5mm}
    \centering 
    \includegraphics[width=\textwidth]{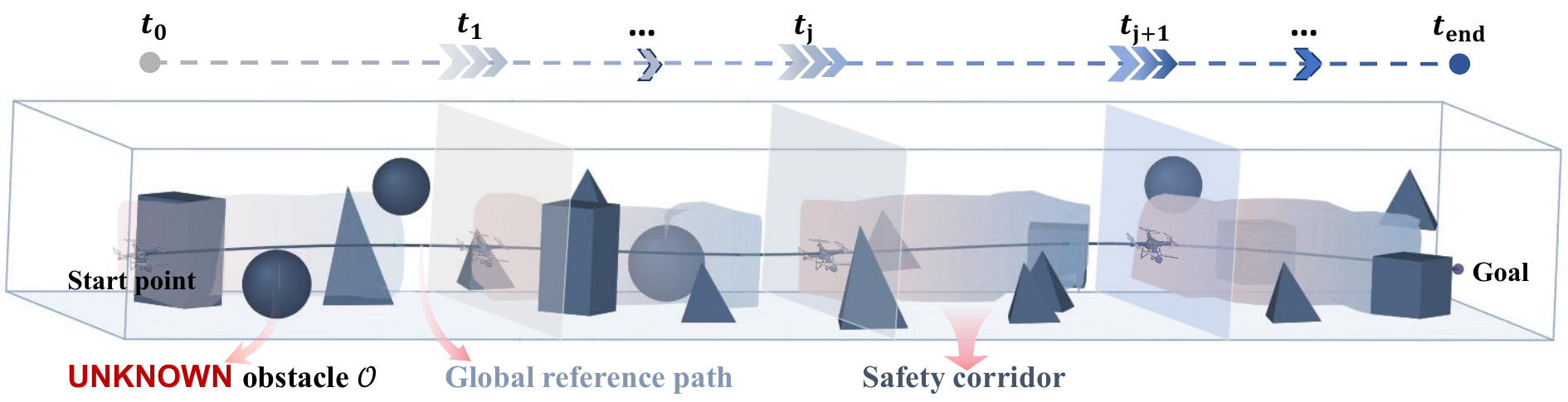}
    \caption{Planning-guided look-ahead safety corridor obtained by fusing the differentiable local SCP functions with adaptive compensation.}
\label{fig:whole corridor}
\end{figure*}

\subsubsection{Corridor Fusion With Adaptive Compensation}
\label{subsubsec:corridor fusion}
Although the local safety function $H_k(p)$ provides a differentiable representation of the obstacle free region around the $k$-th look ahead seed, the spatial coverage of an individual SCP remains limited. In cluttered environments, relying solely on a single SCP may significantly reduce the available traversable space, especially in the presence of elongated obstacles or densely distributed obstacle clusters. The retained local safety functions must therefore be combined to obtain a unified representation of the available motion space along the reference path. In the following fusion step, $N_c$ denotes the number of retained local regions.

One common approach is to use the softmax with a fixed compensation term introduced in Section~\ref{subsec:softmin_softmax}. When applied to the retained local safety functions, it is written as $\frac{1}{\kappa} \ln(\sum_{k=1}^{N_c} e^{\kappa H_k(p)})-\frac{\ln N_c}{\kappa}$ and satisfies  
\begin{equation}
\begin{aligned}
    \max_{k\in\{1,\ldots,N_c\}}H_k(p) - \frac{\ln N_c}{\kappa} 
    &\leq \frac{1}{\kappa} \ln\left(\sum_{k=1}^{N_c} e^{\kappa H_k(p)}\right)
    -\frac{\ln N_c}{\kappa}
    \\
    &\leq \max_{k\in\{1,\ldots,N_c\}}H_k(p).
\end{aligned}
\label{eq:conventional_corridor_fusion_bound}
\end{equation}
The constant compensation $\ln N_c/\kappa$ is independent of the relative contributions of the local safety functions. It corresponds to the largest possible entropy contribution and is therefore determined by the worst case in which all local functions contribute uniformly. In practical navigation, however, the relative contributions of neighboring SCPs vary with the robot position. A single local function generally dominates in the interior of one SCP, whereas several neighboring functions may have similar values near their overlapping regions.

To characterize these location-dependent contributions, we define
\begin{equation*}
    w_k(p)=\frac{e^{\kappa H_k(p)}}{\sum_{j=1}^{N_c}e^{\kappa H_j(p)}}, \qquad k=1,\ldots,N_c.
\end{equation*}
These weights satisfy $w_k(p) > 0$ and $\sum_{k=1}^{N_c} w_k(p) = 1$ and form a probability distribution over the retained local safety functions. A large value of $w_k(p)$ indicates that the $k$th local function has a dominant contribution at $p$, whereas comparable weights indicate that several neighboring local regions contribute simultaneously.

The adaptive compensation is defined from the Shannon entropy of the Softmax weights as
\begin{equation}
    b^*(p) = -\frac{1}{\kappa} \sum_{k=1}^{N_c} w_k(p)\ln w_k(p).
    \label{eq:adaptive_compensation}
\end{equation}
When one or a few local functions dominate, the Softmax distribution is concentrated and $b^*(p)$ becomes small. When the local functions have comparable values, the distribution approaches the uniform distribution and $b^*(p)$ increases toward the conventional compensation $\ln N_c/\kappa$. The compensation therefore adapts to the local geometric contributions rather than applying the same worst case value throughout the entire corridor.

\begin{lemma}[Bound of the Adaptive Compensation]
For every $p\in\mathbb{R}^{d}$, the adaptive compensation defined in~\eqref{eq:adaptive_compensation} satisfies
\begin{equation}
    0 \leq b^*(p) \leq \frac{\ln N_c}{\kappa}.
    \label{eq:adaptive_compensation_bound}
\end{equation}
The upper bound is attained when $w_k(p)=1/N_c$ for all $k\in\{1,\ldots,N_c\}$.
\end{lemma}

\begin{proof}
Since $0<w_k(p)\leq1$, one has
\begin{equation*}
    -w_k(p)\ln w_k(p)\geq0, \qquad k=1,\ldots,N_c,
\end{equation*}
which establishes the lower bound. By the concavity of the logarithm, 
\begin{equation*}
\begin{aligned}
    -\sum_{k=1}^{N_c} w_k(p)\ln w_k(p)
    &\leq \ln(\sum_{k=1}^{N_c} w_k(p) \frac{1}{w_k(p)}) =\ln N_c.
\end{aligned}
\end{equation*}
Dividing the resulting inequality by $\kappa>0$ gives~\eqref{eq:adaptive_compensation_bound}. Equality in the entropy upper bound holds for the uniform distribution $w_k(p)=1/N_c$.
\end{proof}
A direct consequence of~\eqref{eq:adaptive_compensation_bound} is
\begin{equation}
 \frac{1}{\kappa} \ln(\sum_{k=1}^{N_c} e^{\kappa H_k(p)}) 
 - b^*(p)
 \geq 
 \frac{1}{\kappa} \ln(\sum_{k=1}^{N_c} e^{\kappa H_k(p)}) 
 - \frac{\ln N_c}{\kappa}.
\label{eq:adaptive_fixed_compensation_comparison}
\end{equation}

Therefore, the adaptive formulation is pointwise no smaller than the fixed-compensation formulation. Its nonnegative superlevel set consequently contains that obtained using the fixed compensation. Equality holds when the Softmax weights are
uniform, whereas a concentrated weight distribution yields a smaller compensation and reduces unnecessary shrinkage of the admissible region.

The log-sum-exp function admits an exact decomposition in terms of the Softmax weighted average and the entropy of its weights~\cite{Nesterov2005smooth}:
\begin{equation}
\begin{aligned}
    \frac{1}{\kappa} \ln \left(\sum_{k=1}^{N_c} e^{\kappa H_k(p)} \right)
    &=\sum_{k=1}^{N_c}  w_k(p)H_k(p)
    \\
    &- \frac{1}{\kappa} \sum_{k=1}^{N_c} w_k(p)\ln w_k(p).
\end{aligned}
\label{eq:log_sum_exp_entropy_decomposition}
\end{equation}
Substituting the adaptive compensation in~\eqref{eq:adaptive_compensation} into~\eqref{eq:log_sum_exp_entropy_decomposition} gives
\begin{equation}
    \frac{1}{\kappa} \ln \left(\sum_{k=1}^{N_c} e^{\kappa H_k(p)} \right) - b^*(p) = \sum_{k=1}^{N_c} w_k(p)H_k(p).
\label{eq:entropy_compensated_equivalence}
\end{equation}
Accordingly, the unified corridor safety function is defined as
\begin{equation}
    H_{\text{corr}}(p) = \sum_{k=1}^{N_c} w_k(p)H_k(p).
\label{eq:corridor_safety_function}
\end{equation}
Thus, the entropy compensated logarithmic formulation is equivalently expressed as a Softmax weighted average of the retained local safety functions. This equality also shows that $b^*(p)$ is not an independently selected heuristic buffer; it is the exact entropy term associated with the Softmax distribution in the log-sum-exp decomposition.

The proposed look ahead safety corridor is defined by the nonnegative superlevel set:
\begin{equation}
    \mathcal{C}_{\text{corr}} = \big\{p\in\mathbb{R}^{d} \mid H_{\text{corr}}(p)\geq0 \big\}.
    \label{eq:corridor_safe_set}
\end{equation}

The overlap condition in~\eqref{eq:local_region_overlap} applies to the local regions before fusion. Since fusion gives an inner approximation of their union, the resulting corridor is checked before use. Let $\Gamma_k^e$ denote the continuous reference
segment associated with the first $k$ seeds, including its connection from the current position $p$. For $\Gamma^e=\Gamma_{N_c}^e$ and a field margin $\varepsilon_H>0$, the acceptance conditions are
\begin{equation}
    \begin{aligned}
        H_{\mathrm{corr}}(p)
        &\geq \varepsilon_H,\\
        \min_{q\in\Gamma^e} H_{\mathrm{corr}}(q)
        &\geq \varepsilon_H,\\
        \operatorname{Int}(\mathcal C_{\mathrm{corr}})
        &\text{ is connected}.
    \end{aligned}
    \label{eq:corridor_acceptance}
\end{equation}
Here, $\varepsilon_H$ is a margin in field values, not a geometric clearance.

If these conditions are not met, the seed spacing is halved and the local regions are reconstructed over the same look-ahead reference. The fused corridor is then checked again. This refinement is repeated up to $M_{\mathrm{ref}}$ times. If the corridor still fails the acceptance test, the last retained region is removed and the reference segment is shortened accordingly. The fusion weights and corridor function are recomputed after each removal. The procedure returns an accepted corridor or an empty output if no candidate is accepted.
Algorithm~\ref{alg:corridor_construction} summarizes the procedure. An example of the resulting corridor is shown in Fig.~\ref{fig:whole corridor}.

\begin{algorithm}[h]
    \caption{Planning-Guided Look-Ahead Safety Corridor Construction}
    \label{alg:corridor_construction}
    \begin{algorithmic}[1]
        \STATE \textbf{Input: }{current collision-free position $p\in\mathbb R^d$; ordered global reference path $\Gamma=\{\gamma_1,\ldots,\gamma_L\}$; 
        obstacle point set $\mathcal{O}$; 
        construction domain $\mathcal D(\cdot)$; 
        corresponding domain-facet function set $\mathcal F_{\mathcal D}$; 
        safety clearance parameter $\rho>0$; 
        smoothing parameter $\kappa>0$};
        field margin $\varepsilon_H>0$;
        look-ahead length $L_a$ and initial seed spacing $\Delta s$;
        maximum refinement count $M_{\mathrm{ref}}$
        
        \STATE \textbf{Output: }{retained SCP sequence $\{\mathcal P_k\}_{k=1}^{N_c}$ and its facet function set $\{F_k\}_{k=1}^{N_c}$;
        local safety functions $\{H_k\}_{k=1}^{N_c}$; 
        retained reference segment $\Gamma^e$;
        corridor function $H_{\mathrm{corr}}$ and corridor $\mathcal C_{\mathrm{corr}}$};

         \FOR{$m=0,\ldots,M_{\mathrm{ref}}$}
            \STATE Generate ordered seeds
            $\{s_k\}_{k=1}^{N}$ with $s_1=p$
            over the look-ahead reference using spacing
            $\Delta s/2^m$, and define their
            reference prefixes $\{\Gamma_k^e\}_{k=1}^{N}$

            \STATE Initialize $N_c\leftarrow0$

            \FOR{$k=1,\ldots,N$}
                \STATE Select
                $\mathcal O_{s_k}\leftarrow
                \{o\in\mathcal O\mid
                \operatorname{dist}(o,\mathcal D(s_k))
                \leq\rho\}$

                \STATE Construct
                $\mathcal P_k$, $\mathcal F_k$, and $H_k$
                using the local SCP construction
                and smooth minimum

                \IF{the local SCP construction is infeasible}
                    \STATE \textbf{break}
                \ENDIF

                \STATE Set
                $\widetilde{\mathcal P}_k
                \leftarrow\{q\mid H_k(q)\geq0\}$

                \IF{$k>1$ and
                    $\operatorname{Int}
                    (\widetilde{\mathcal P}_{k-1})
                    \cap
                    \operatorname{Int}
                    (\widetilde{\mathcal P}_k)
                    =\varnothing$}
                    \STATE \textbf{break}
                \ENDIF

                \STATE $N_c\leftarrow k$
            \ENDFOR

            \WHILE{$N_c\geq1$}
                \STATE Set
                $\Gamma^e\leftarrow\Gamma_{N_c}^e$

                \STATE Recompute the fusion weights
                and $H_{\mathrm{corr}}$ using
                $\{H_k\}_{k=1}^{N_c}$

                \STATE Set
                $\mathcal C_{\mathrm{corr}}
                \leftarrow
                \{q\mid H_{\mathrm{corr}}(q)\geq0\}$

                \IF{the conditions
                    in~\eqref{eq:corridor_acceptance}
                    hold}
                    \RETURN
                    $\{\mathcal P_k,\mathcal F_k,H_k\}_{k=1}^{N_c}$,
                    $\Gamma^e$,
                    $H_{\mathrm{corr}}$, and
                    $\mathcal C_{\mathrm{corr}}$
                \ENDIF

                \IF{$m<M_{\mathrm{ref}}$}
                    \STATE \textbf{break}
                \ENDIF

                \STATE $N_c\leftarrow N_c-1$
            \ENDWHILE
        \ENDFOR

        \RETURN $\varnothing$
    \end{algorithmic}
\end{algorithm}

The following proposition establishes the geometric safety property required for the subsequent Barrier State embedding.
\begin{proposition}[Safety Set Containment]
\label{prop:corridor_safeset}
For every $p\in\mathbb{R}^{d}$, the unified corridor safety function in~\eqref{eq:corridor_safety_function} satisfies
\begin{equation}
    H_{\text{corr}}(p) \leq \max_{k\in\{1,\ldots,N_c\}} H_k(p).
    \label{eq:corridor_maximum_bound}
\end{equation}
Consequently, the corridor safe set satisfies
\begin{equation*}
    \mathcal{C}_{\text{corr}} \subseteq \bigcup_{k=1}^{N_c} \widetilde{\mathcal{P}}_k
    \subseteq \bigcup_{k=1}^{N_c} \mathcal{P}_k.
    \label{eq:corridor_safe_set_containment}
\end{equation*}
\end{proposition}
\begin{proof}
    The Softmax weights satisfy $w_k(p)>0$ and $\sum_{k=1}^{N_c}w_k(p)=1$. Therefore,
\begin{equation*}
\begin{aligned}
    H_{\mathrm{corr}}(p)=\sum_{k=1}^{N_c} w_k(p)H_k(p)
    &\leq \sum_{k=1}^{N_c} w_k(p) \max_{j\in\{1,\ldots,N_c\}} H_j(p)
    \\
    &=\max_{j\in\{1,\ldots,N_c\}} H_j(p),
\end{aligned}
\end{equation*}
which proves~\eqref{eq:corridor_maximum_bound}. If $H_{\mathrm{corr}}(p)\geq0$, then~\eqref{eq:corridor_maximum_bound} implies
$\max_{k\in\{1,\ldots,N_c\}}H_k(p)\geq0$. Hence, there exists at least one $k\in\{1,\ldots,N_c\}$ such that $H_k(p)\geq0$. According to
\eqref{eq:smooth_local_safe_region}, this implies $p\in\widetilde{\mathcal{P}}_k$.
\end{proof}

Proposition~\ref{prop:corridor_safeset} establishes containment within the local safe-region union.
Together with the adaptive compensation bound, it shows that,
for the same local functions and smoothing parameter,
the proposed safe set contains the fixed-compensation approximation
without extending beyond that union.

Unlike the temporal composition of recently perceived safe sets~\cite{safari2025time}, the local regions here are organized along
a forward reference path.
The acceptance conditions additionally check the current position
and retained path in the final fused region.

\subsection{Poisson Safety Field Construction}
\label{subsec:Poisson}
Although $H_{\mathrm{corr}}$ provides a differentiable representation of the connected safety corridor, its interior values still inherit the local geometry of the fused SCPs. Consequently, the magnitude and direction of $\nabla H_{\mathrm{corr}}$ may vary near the transition between adjacent SCPs. To obtain a spatially consistent interior field without changing the corridor boundary, a Poisson safety field is constructed over the corridor domain. This construction follows the use of Poisson equations for synthesizing smooth safety functions from geometric boundaries~\cite{bahati2025dynamic}.

For a corridor accepted by Algorithm~\ref{alg:corridor_construction}, let
\begin{equation}
    \Omega_{\mathrm{corr}} :=
    \operatorname{Int}(\mathcal C_{\mathrm{corr}}).
\end{equation}
denote its connected interior. The acceptance conditions ensure that this domain is connected and contains the current position and retained reference segment. The following Poisson construction additionally requires the boundary regularity stated below:
\begin{equation*}
    \partial\Omega_{\mathrm{corr}} =
    \left\{
    p\in\mathbb{R}^{d}
    \mid
    H_{\mathrm{corr}}(p)=0
    \right\}.
\end{equation*}

The Poisson safety field $H_{\mathrm{P}}$ is defined as the solution of
\begin{equation}
\label{eq:poisson_safety_field}
\begin{cases}
    - \Delta H_{\mathrm{P}}(p)=f_P(p),
    & p\in\Omega_{\mathrm{corr}},\\[2mm]
    H_{\mathrm{P}}(p)=0,
    & p\in\partial\Omega_{\mathrm{corr}},
\end{cases}
\end{equation}
where $\Delta$ is the Laplace operator. For example, when $d=3$, $\Delta= \dfrac{\partial^{2}}{\partial p_{x}^{2}} + \dfrac{\partial^{2}}{\partial p_{y}^{2}} + \dfrac{\partial^{2}}{\partial p_{z}^{2}}$. The source function satisfies
\begin{equation}
    0<f_{\min}\le f_P(p)\le f_{\max}<\infty
    \quad\text{a.e. in }\Omega_{\mathrm{corr}}.
    \label{eq:poisson_source_bounds}
\end{equation}
The source controls the interior field distribution, while the zero Dirichlet condition specifies the corridor boundary. The constant-source construction is recovered by setting $f_P\equiv1$.

Assume that zero is a regular value of $H_{\mathrm{corr}}$ on the selected corridor boundary and that $\Omega_{\mathrm{corr}}$ is bounded with a $C^2$ boundary. For $f_P\in L^\infty(\Omega_{\mathrm{corr}})$, the Poisson problem admits a unique weak solution. Let $r\in(1,\infty)$ be a real-valued exponent that specifies the integrability of the solution and its weak derivatives. The space $W^{2,r}(\Omega_{\mathrm{corr}})$ consists of functions whose weak derivatives up to order two, including the functions themselves, belong to $L^r(\Omega_{\mathrm{corr}})$. Elliptic regularity gives $H_P\in W^{2,r}(\Omega_{\mathrm{corr}})$ for every finite $1<r<\infty$~\cite{gilbarg2001elliptic}. Taking $r>d$ yields a continuously differentiable representative on the corridor closure, where $d$ is the workspace dimension. Thus, the field provides the continuous first-order spatial derivatives required for BaS embedding, even when the source is only essentially bounded. These regularity conditions are assumptions on the accepted corridor domain and are not guaranteed by the fusion construction alone. Under~\eqref{eq:poisson_source_bounds}, the strong maximum principle gives
\begin{equation*}
\label{eq:poisson_positive}
    H_{\mathrm{P}}(p)>0,
    \qquad
    p\in\Omega_{\mathrm{corr}}.
\end{equation*}
Together with the zero boundary condition, this preserves the prescribed corridor interior and boundary. 

Let $n(p)$ denote the outward unit normal to the corridor boundary. Hopf's boundary-point lemma further yields~\cite{gilbarg2001elliptic}
\begin{equation}
\label{eq:poisson_boundary_gradient}
    \nabla H_{\mathrm{P}}(p)^{\top}n(p)<0,
    \qquad
    p\in\partial\Omega_{\mathrm{corr}}.
\end{equation}
Hence, the field gradient points toward the interior at the boundary. The interior field depends on both the corridor boundary and the prescribed source, rather than only on individual SCP facets.

\begin{figure}[!tbp]
\vspace{5mm}
    \centering 
    \includegraphics[width=\columnwidth]{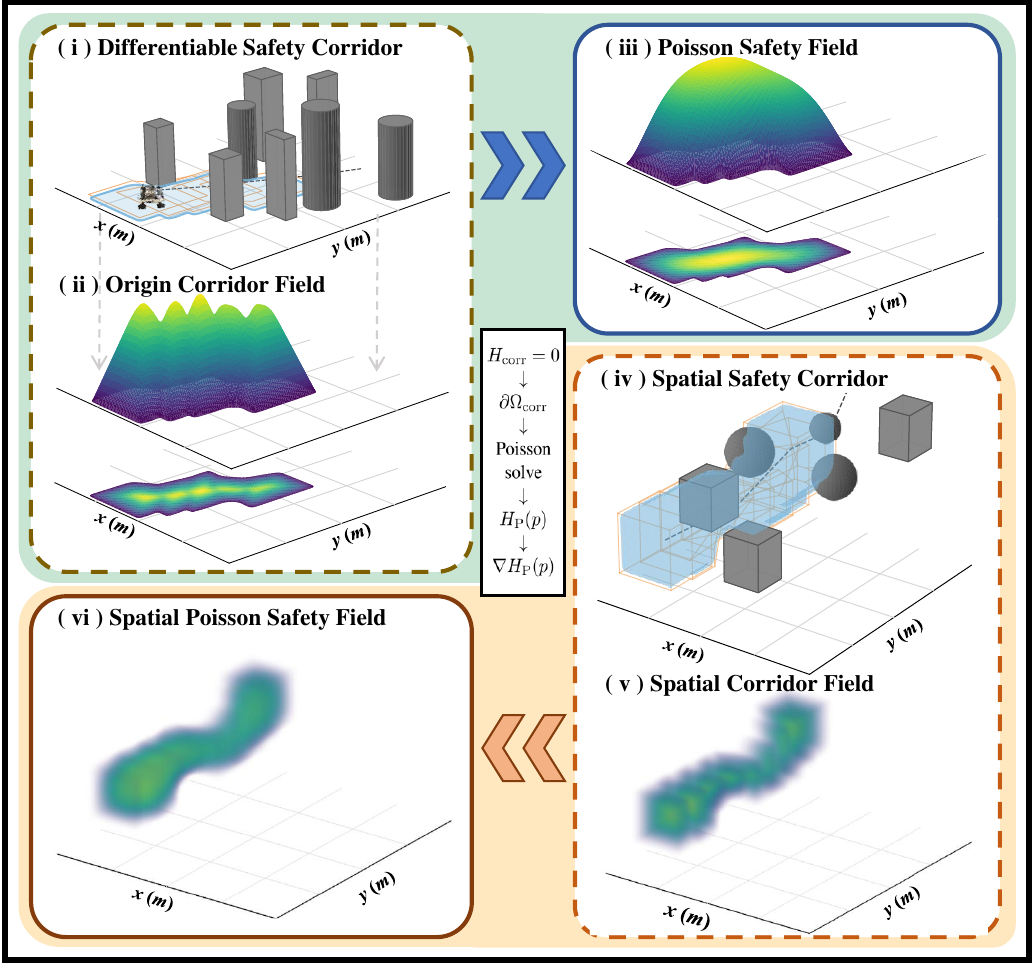}
    \caption{Poisson safety field construction for planar and
spatial corridors. (i) Planar safety corridor. (ii) Original corridor field $H_{\mathrm{corr}}$. (iii) Reconstructed Poisson field $H_{\mathrm{P}}$. (iv) Spatial safety corridor. (v) Original spatial corridor field. (vi) Reconstructed spatial Poisson field. The reconstruction retains the prescribed boundary and redistributes the interior field according to the Poisson problem.}
\label{fig:poisson}
\end{figure}

Fig.~\ref{fig:poisson} illustrates the Poisson reconstruction for planar and spatial corridors. The prescribed boundary is retained, while the interior field is determined by the source and boundary conditions. The resulting field is used for BaS embedding.

\section{Framework of Barrier-State-Based Safe Navigation}
\label{sec:framwork}
Section IV constructs a differentiable safety corridor and defines the Poisson field $H_{\mathrm{P}}$ over its interior. In this section, we embed the field into the
Barrier-State (BaS) framework and establish a conditional
safety result for the active corridor. The analysis follows
the BaS consistency and boundedness results in~\cite{almubarak2025barrier}. It does not require a specific controller structure. Instead, the selected feedback must satisfy the input constraints and keep the barrier state bounded.

\subsection{Barrier State Embedding}
\label{subsec:bas embedding}
For each accepted corridor, let
$\Omega_{\mathrm{corr}}=\operatorname{Int}
(\mathcal{C}_{\mathrm{corr}})$.
The Poisson field is continuously extended to
$\overline{\Omega}_{\mathrm{corr}}$.
Define
\begin{equation*}
h_{\mathrm{corr}}(x) = H_P\bigl(\pi(x)\bigr),
\qquad
\pi(x)\in\overline{\Omega}_{\mathrm{corr}}.
\end{equation*}
Under the regularity conditions in Section~\ref{subsec:Poisson} and a continuously differentiable position map $\pi$, this function is continuously differentiable in the corridor interior. Moreover,
\begin{equation*}
h_{\mathrm{corr}}(x)>0
\quad \text{if } \pi(x)\in\Omega_{\mathrm{corr}},
\end{equation*}
and
\begin{equation*}
h_{\mathrm{corr}}(x)=0
\quad \text{if } \pi(x)\in\partial\Omega_{\mathrm{corr}}.
\end{equation*}

Let $B$ be the barrier transformation defined in
Section~\ref{subsec:cbf}. For a fixed finite offset $\beta_e$, define
\begin{equation}
z=B\bigl(h_{\mathrm{corr}}(x)\bigr)-\beta_e.
\label{eq:corridor_bas}
\end{equation}

If an admissible equilibrium $x_e$ lies inside the active
corridor, choosing $\beta_e=B(h_{\mathrm{corr}}(x_e))$ gives $z=0$ at $x_e$.  The safety result below only requires $\beta_e$ to be finite and fixed; it does not require convergence to an equilibrium.

The state gradient follows from the position mapping:
\begin{equation*}
    \nabla_x h_{\mathrm{corr}}(x) = J_{\pi}(x)^{\mathsf T}
    \nabla_p H_{\mathrm{P}}\bigl(\pi(x)\bigr),
    \label{eq:poisson_state_gradient}
\end{equation*}
where$J_{\pi}(x)=\partial\pi(x)/\partial x \in\mathbb{R}^{d\times n}$. Consequently,
\begin{equation}
    \begin{aligned}
        L_f h_{\mathrm{corr}}(x)
        &=
        \nabla_p H_{\mathrm{P}}\bigl(\pi(x)\bigr)^{\mathsf T} J_{\pi}(x)f(x),\\
        L_g h_{\mathrm{corr}}(x)
        &=
        \nabla_p H_{\mathrm{P}}\bigl(\pi(x)\bigr)^{\mathsf T} J_{\pi}(x)g(x).
    \end{aligned}
    \label{eq:poisson_lie_derivatives}
\end{equation}

The active field is held fixed between corridor updates. With consistent initialization, the BaS consistency result
in~\cite{almubarak2025barrier} gives
\begin{equation*}
z(t)=B\bigl(h_{\mathrm{corr}}(x(t))\bigr)-\beta_e.
\end{equation*}
Hence, along the nominal system,
\begin{equation*}
\dot z
=
B'\bigl(h_{\mathrm{corr}}(x)\bigr)
\left(
L_fh_{\mathrm{corr}}(x)
+
L_gh_{\mathrm{corr}}(x)u
\right).
\end{equation*}
The augmented state is $X=[x^\top,z]^\top$. Its dimension is $n+1$, independently of the number of obstacle points, facets, or local SCPs. When the state and field are available, $z$ can be evaluated directly from~\eqref{eq:corridor_bas}.

\subsection{Safety Feedback and Corridor Invariance}
The accepted corridor and its Poisson field provide the geometric constraint for BaS-based control. Using the consistency and boundedness results in~\cite{almubarak2025barrier}, the following theorem establishes safety within an active corridor, provided that the corridor remains collision-free and the selected
admissible feedback keeps the barrier state bounded.

\begin{theorem}[Safety within the active corridor]
\label{thm:poisson_corridor_safety}
Consider an update interval $[t_j,t_{j+1})$. Suppose that:
\begin{enumerate}
    \item The active corridor is fixed and remains
    collision-free throughout the interval, with the
    prescribed clearance accounting for the robot extent and the assumed geometric uncertainty.
    \item Its Poisson field satisfies the regularity,
    positivity, and boundary conditions in Section~\ref{subsec:Poisson}.
    \item The initial position satisfies $\pi(x(t_j))\in\Omega_{\mathrm{corr},j}$, and the
    barrier state is consistently initialized as
    \begin{equation*}
    z(t_j)
    =
    B\bigl(h_{\mathrm{corr},j}(x(t_j))\bigr)-\beta_e,
    \end{equation*}
    where $\beta_e$ is finite and fixed.
\end{enumerate}
If an admissible feedback $u(t)\in\mathcal U$ yields a
well-posed closed-loop solution with no finite-time escape and keeps $z$ uniformly bounded above along its interior evolution on this interval, then
\begin{equation}
    h_{\mathrm{corr},j}(x(t))>0,\qquad
    \pi(x(t))\in\Omega_{\mathrm{corr},j},
    \quad t\in[t_j,t_{j+1}).
    \label{eq:poisson_corridor_invariance}
\end{equation}
Consequently, the robot remains within the active collision-free corridor.

\end{theorem}
\begin{proof}
The consistent initialization and the barrier-state dynamics preserve
\begin{equation}
    B\bigl(h_{\mathrm{corr},j}(x(t))\bigr)= z(t)+\beta_e,
    \label{eq:poisson_barrier_consistency}
\end{equation}
as long as the position remains in the corridor interior. Since $z$ is bounded above and $\beta_e$ is fixed and finite, the right-hand side of~\eqref{eq:poisson_barrier_consistency} has a finite upper bound.

Suppose that the position first reaches the corridor boundary at a finite time $t^\ast \in(t_j,t_{j+1})$. Continuity of the trajectory and the boundary extension of $H_{P,j}$ gives
\begin{equation*}
h_{\mathrm{corr},j}(x(t))\rightarrow0^+
\quad\text{as }t\rightarrow t^{\ast-}.
\end{equation*}
The defining property of $B$ then implies
\begin{equation*}
B\bigl(h_{\mathrm{corr},j}(x(t))\bigr) \rightarrow+\infty,
\end{equation*}
which contradicts its finite upper bound. Hence, the position cannot reach the corridor boundary during the active interval. The assumed collision-free validity of the active corridor then establishes physical safety.

By Proposition~\ref{prop:corridor_safeset}
in Section~\ref{subsubsec:corridor fusion}, the active corridor satisfies
\begin{equation}
    \pi(x(t))
    \in\Omega_{\mathrm{corr},j}
    \subseteq\mathcal{C}_{\mathrm{corr},j}
    \subseteq\bigcup_{k=1}^{N_{c,j}}\widetilde P_{k,j}
    \subseteq\bigcup_{k=1}^{N_{c,j}}P_{k,j},
    \label{eq:poisson_geometric_containment}
\end{equation}
where $N_{c,j}$ is the number of local regions
retained at update $j$. Together with the collision-free condition on the active corridor, this establishes the claim.
\end{proof}

% The boundedness condition in Theorem~\ref{thm:poisson_corridor_safety} must be established for the selected control design. One sufficient certificate is a nonnegative, continuously differentiable function $V(X)$ satisfying
% \begin{equation}
%     \alpha(|z|) \leq V(X), \hspace{3mm} \nabla_X V(X)^\top\bigl(F(X)+G(X)u\bigr) \leq 0,
%     \label{eq:augmented_boundedness_certificate}
% \end{equation}
% along the closed-loop solution, where $\alpha$ is a class-$\mathcal{K}_\infty$ function. These inequalities imply
% \begin{equation*}
%      |z(t)| \leq \alpha^{-1}\bigl(V(X(t_j))\bigr).
% \end{equation*}
For $B(h)=1/h$, an upper bound $z\leq M_z$ gives
\begin{equation}
    h_{\mathrm{corr},j}(x(t))
    \geq\frac{1}{M_z+\beta_e}>0.
\end{equation}
This is a margin in field values, not a Euclidean clearance.

The following corollary gives a sufficient controller
compatibility condition based on BaS theory~\cite{almubarak2025barrier}.

\begin{corollary}
\label{cor:bas_compatibility}

Consider a frozen-corridor mode satisfying the geometric and initialization assumptions of
Theorem~\ref{thm:poisson_corridor_safety}. Let $\bar X$ denote the full closed-loop state, containing $X=[x^\top,z]^\top$ and any additional controller, actuator, or barrier states. Let $\bar X_{e,j}$ be a consistent equilibrium of the undisturbed mode. Suppose that the selected feedback respects the input constraints and makes the full closed-loop system locally input-to-state stable (ISS) with respect to an admissible bounded input disturbance $d_{\mathrm{ext}}$~\cite{mironchenko2016local}. For initial states $\bar X(t_j^+)\in\mathcal A_j$ and disturbances within the certified bounds, there exist comparison functions
$\alpha_j\in\mathcal{KL}$ and $\mu_j\in\mathcal K$ such that
\begin{equation}
\begin{aligned}
    &\|\bar X(t)-\bar X_{e,j}\|\\
    &\quad\leq
    \alpha_j\!\left(
        \|\bar X(t_j^+)-\bar X_{e,j}\|,\,
        t-t_j
    \right)\\
    &\qquad+
    \mu_j\!\left(
        \|d_{\mathrm{ext}}\|_{\infty,[t_j,t]}
    \right),
    \qquad t\in[t_j,t_{j+1}).
\end{aligned}
\label{eq:closed_loop_iss}
\end{equation}

Here, $\mathcal A_j$ is the certified initial-state domain, and the disturbance norm is its essential supremum over the indicated time interval. The class-$\mathcal{KL}$ term decays with elapsed time, whereas the class-$\mathcal K$ term bounds the disturbance effect. Assume that BaS consistency is preserved for the perturbed dynamics.

Then the original state error $x-x_{e,j}$ is locally ISS, where $x_{e,j}$ is the physical-state component of $\bar X_{e,j}$. The geometric constraint is locally ISSf in the sense of~\cite{almubarak2025barrier}. Thus, the frozen-corridor system is locally input-to-state safely stable ($\mathrm{IS}^{3}$). The robot remains inside the corridor while that corridor is valid and active.

\end{corollary}

\begin{proof}
Let $z_{e,j}$ be the geometric barrier-state component
of $\bar X_{e,j}$.

Since $z$ is a component of $\bar X$,
\begin{equation*}
    |z(t)|
    \leq
    |z_{e,j}|+\|\bar X(t)-\bar X_{e,j}\|.
\end{equation*}

Estimate~\eqref{eq:closed_loop_iss} therefore bounds $z$ for the certified initial states and disturbances. BaS consistency then bounds $B(h_{\mathrm{corr},j}(x(t)))$. The boundary argument in
Theorem~\ref{thm:poisson_corridor_safety} gives corridor invariance. The corresponding local ISS and ISSf properties follow from the BaS result~\cite{almubarak2025barrier}.

\end{proof}

A global conclusion requires a global ISS certificate over the admissible consistent state space. If the corridor remains fixed and valid for all future time, zero disturbance gives asymptotic convergence to the certified equilibrium. Bounded nonzero disturbances give an ISS bound, without guaranteeing exact convergence. Existing control designs, including a suitable PID realization, may be used when the certificate covers the full closed loop and its reference dynamics.

\begin{remark}
At a corridor activation time $t_j$, the candidate must satisfy the geometric acceptance conditions. In particular,
\begin{equation}
    H_{\mathrm{corr},j}
    \bigl(\pi(x(t_j))\bigr)>0.
    \label{eq:corridor_update_admission}
\end{equation}

The physical state remains continuous, and the geometric barrier state is reset or directly reevaluated as
\begin{equation*}
    \begin{aligned}
        x(t_j^+)&=x(t_j^-),\\
        z(t_j^+)&=
        B\!\left(
            H_{P,j}\bigl(\pi(x(t_j))\bigr)
        \right)-\beta_e.
    \end{aligned}
\end{equation*}

When Corollary~\ref{cor:bas_compatibility} is invoked, the full state after the update must satisfy $\bar X(t_j^+)\in\mathcal A_j$. Retained controller and actuator states must be included in this check, and any additional reset must be covered by the certificate.

Theorem~\ref{thm:poisson_corridor_safety} applies successively
if its assumptions hold on every active interval and the update times have no finite accumulation point. A rejected candidate is not activated. The previous corridor can be retained only while it remains valid and the required boundedness condition holds. Safety on successive intervals does not by itself establish ISS of the switching closed loop or convergence to the final navigation goal.
\end{remark}

\section{SIMULATIONS AND EXPERIMENTS}
\label{sec:sim_and_car}
\begin{figure*}[!t]
    \centering
    \includegraphics[width=\textwidth]{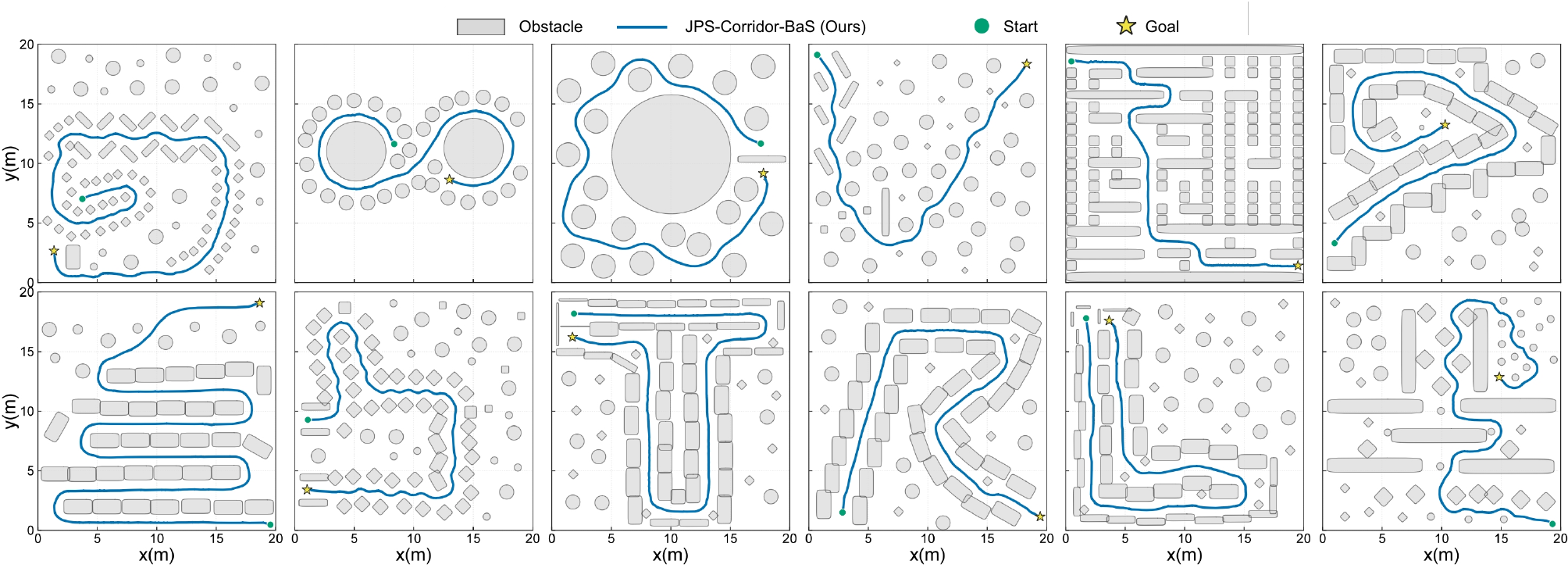}
    \caption{Representative navigation results of the proposed method in S1. The panels show final executed trajectories in 12 selected environments. Gray regions denote obstacles, the gray dashed curves denote the final valid JPS paths, and the solid curves denote the executed robot trajectories. The green circles and yellow stars indicate the start and goal positions, respectively.}
    \label{fig:s1_run}
\end{figure*}
\begin{figure*}[!t]
    \centering
    \includegraphics[width=\textwidth]
    {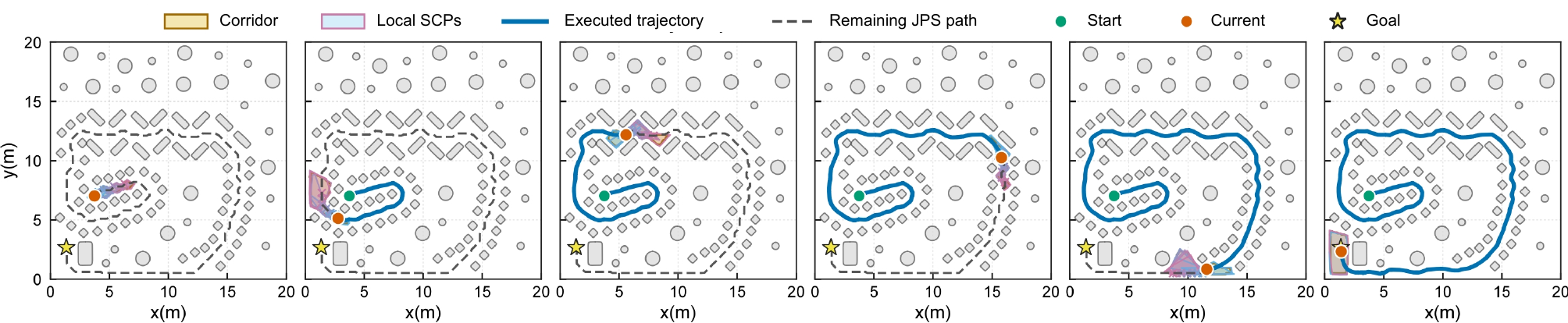}
    \caption{Process diagram of the proposed method in a representative S1 environment.}
    \label{fig:s1_process}
\end{figure*}

\begin{figure*}[!t]
    \centering
    \includegraphics[width=\textwidth]{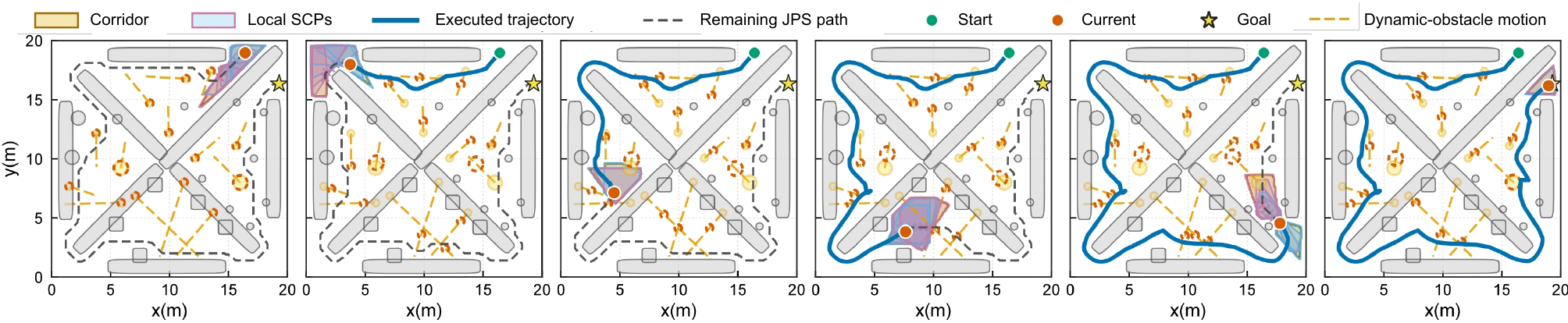}
    \caption{Process diagram of the proposed method in a representative S2 environment. The local SCPs and their fused corridor are updated as the robot moves and the free space changes. The blue curve denotes the executed trajectory, the dashed curve denotes the remaining valid JPS path, and the orange marker denotes the current robot position.}
    \label{fig:s2_process}
\end{figure*}

\subsection{Safe Navigation in Complex Environments (Simulations)}

We evaluate the proposed method in a $20\,\mathrm{m}\times20\,\mathrm{m}$ workspace.
The robot radius is $0.28\,\mathrm{m}$, the additional safety margin is $0.10\,\mathrm{m}$, and the workspace boundaries are physical walls. Two sets are considered: S1 contains static obstacles, whereas S2 includes moving obstacles. Each set contains 25 environments. 

The robot position satisfies $\dot p=u$, where $p,u\in\mathbb R^2$ are the position and applied velocity. We introduce the dynamic extension $\dot u=\nu$ to handle the speed constraint $\|u\|<\bar u$. For the active geometric safety field $h(p)$, the geometric and input barrier states are
\begin{equation}
\label{eq:sim_barrier_states}
\begin{aligned}
    z_c&=\frac{1}{h(p)}-\beta_e,\\
    z_u&=\frac{1}{h_{\mathrm{in}}(u)}-1,
    \qquad
    h_{\mathrm{in}}(u)=1-\frac{u^\top u}{\bar u^2}.
\end{aligned}
\end{equation}
These definitions require $h(p)>0$ and $h_{\mathrm{in}}(u)>0$.

Let $X=[p^\top,u^\top,z_c,z_u]^\top$. During each interval with a fixed safety field
and offset, the consistent augmented dynamics are
\begin{equation}
\begin{aligned}
    \dot X
    &=
    \underbrace{
    \begin{bmatrix}
        0&I_2&0&0\\
        0&0&0&0\\
        0&\eta_c^\top&0&0\\
        0&0&0&0
    \end{bmatrix}}_{\mathcal A_{\mathrm{sim}}(X)}
    X
    +
    \underbrace{
    \begin{bmatrix}
        0\\I_2\\0\\\eta_u^\top
    \end{bmatrix}}_{\mathcal B_{\mathrm{sim}}(X)}
    \nu,\\
    u&=
    \underbrace{
    \begin{bmatrix}
        0&I_2&0&0
    \end{bmatrix}}_{C_u}X,
\end{aligned}
\label{eq:sim_matrix_model}
\end{equation}
where the zero blocks have compatible dimensions and
\begin{equation}
    \eta_c=-\frac{\nabla_p h(p)}{h(p)^2},
    \qquad
    \eta_u=\frac{2u}{\bar u^2h_{\mathrm{in}}(u)^2}.
\end{equation}
The matrices are state dependent. At each field update, $z_c$ is reevaluated, while $u$ and $z_u$ are retained.

The controller is
\begin{equation}
\begin{aligned}
    u_{\mathrm{des}}
        &=k_p(p_{\mathrm{ref}}-p)+u_{\mathrm{safe}},\\
    \nu
        &=\lambda(u_{\mathrm{des}}-u)
          -\kappa_u z_u\eta_u,
\end{aligned}
\label{eq:sim_control_law}
\end{equation}
where $p_{\mathrm{ref}}$ is the local reference position. The input barrier feedback limits the applied velocity through the dynamic extension. The robot receives $u=C_uX$, rather than $\nu$.

For numerical control evaluation, define
\begin{equation}
    \widehat z_c
        =\frac{1}{\max\{h(p),\varepsilon\}}-\beta_e,
    \qquad
    r_c=k_c[\widehat z_c]_+\nabla_p h(p),
\end{equation}
where $[a]_+=\max\{a,0\}$. For the BaS-based methods, $h$ is the corresponding geometric safety field, and
\begin{equation}
    u_{\mathrm{safe}}=
    \begin{cases}
        \displaystyle
        \frac{r_c}{\max\{1,\|r_c\|/r_{\max}\}},
        & \text{S1},\\[1ex]
        r_c, & \text{S2}.
    \end{cases}
\end{equation}
JPS uses $u_{\mathrm{safe}}=0$. The regularized variable $\widehat z_c$ is used for numerical evaluation; its finite value does not establish boundedness of the ideal state $z_c$.

For the proposed method, forward navigation uses the unnormalized Poisson safety field with
\begin{equation}
    f_P(p)=1-\delta_P\sigma(p),
    \qquad \delta_P=0.10,
    \label{eq:sim_poisson_source}
\end{equation}
where $\sigma(p)\in[0,1]$ represents progress along the forward reference path. For each grid point, the nearest sampled reference point is selected, and its normalized arc-length progress defines $\sigma(p)$. Thus, $0.9\le f_P(p)\le1$. The source remains fixed during each active-field interval. The sampled progress can be discontinuous. The field is therefore interpreted through the weak Poisson formulation. The implementation uses a finite-difference solve with a grid spacing of $0.10\,\mathrm{m}$ and bilinear interpolation. The resulting numerical field approximates the continuous solution and its boundary.

The parameters are $\bar u=5\,\mathrm{m/s}$, $k_p=1.2\,\mathrm{s}^{-1}$, $k_c=0.35$,
$\lambda=200\,\mathrm{s}^{-1}$, $\kappa_u=0.02$, $\varepsilon=10^{-3}$, and
$r_{\max}=1.2\,\mathrm{m/s}$. The sampling period is $0.02\,\mathrm{s}$. Velocity is updated by implicit Euler within the admissible input set, and position is updated
by trapezoidal integration. The maximum path look-ahead distance is $3.0\,\mathrm{m}$, the maximum seed spacing is $0.5\,\mathrm{m}$, and the softmin and softmax constructions both use $\kappa=15$.

We compare JPS, JPS-BaS, JPS-SCP-BaS, and the proposed JPS-Corridor-BaS, giving 200 runs over 50 environments. All methods share the environment layouts, start and goal positions, robot dimensions, input-constraint controller, and termination criteria. A run succeeds if the robot enters the goal region of radius $0.35\,\mathrm{m}$ within $60\,\mathrm{s}$without collision. Collision checks account for the full robot footprint and the workspace walls. Failures are classified as collision (C), deadlock (D), or timeout (T). Navigation time is the simulated start-to-goal time,
averaged over successful runs only.

\subsubsection{Static Environments}
\begin{table}[!tbp]
    \centering
    \scriptsize
    \setlength{\tabcolsep}{3pt}
    \renewcommand{\arraystretch}{1.15}
    \caption{Navigation outcomes in the selected S1 environments.}
    \label{tab:static_outcomes}
    
    \begin{tabular*}{\columnwidth}{
        @{\extracolsep{\fill}}cccc>{\columncolor{gray!10}}c@{}
    }
    \toprule
    \diagbox[
        width=0.25\columnwidth,
        height=0.85cm,
        trim=l,
        linewidth=0.4pt,
        font=\scriptsize
    ]{\textbf{Environment}}{\textbf{Method}}
        & JPS
        & JPS--BaS
        & \shortstack{JPS--SCP--BaS}
        & \shortstack{\textbf{JPS--Corridor--BaS} \\\textbf{(Ours)}}
        \\
        \midrule
        1  & \ResultFail{C} & \ResultFail{D}
           & \ResultPass   & \ResultPass \\
        2  & \ResultPass   & \ResultPass
           & \ResultPass   & \ResultPass \\
        3  & \ResultPass   & \ResultPass
           & \ResultPass   & \ResultPass \\
        4  & \ResultPass   & \ResultPass
           & \ResultPass   & \ResultPass \\
        5  & \ResultFail{C} & \ResultFail{D}
           & \ResultPass   & \ResultPass \\
        6  & \ResultPass   & \ResultFail{D}
           & \ResultPass   & \ResultPass \\
        7  & \ResultFail{C} & \ResultFail{D}
           & \ResultFail{T} & \ResultPass \\
        8  & \ResultFail{C} & \ResultFail{D}
           & \ResultPass   & \ResultPass \\
        9  & \ResultFail{C} & \ResultFail{D}
           & \ResultPass   & \ResultPass \\
        10 & \ResultPass   & \ResultFail{D}
           & \ResultPass   & \ResultPass \\
        11 & \ResultFail{C} & \ResultFail{D}
           & \ResultPass   & \ResultPass \\
        12 & \ResultFail{C} & \ResultPass
           & \ResultPass   & \ResultPass \\
        \bottomrule
    \end{tabular*}
    \vspace{3pt}
    \begin{minipage}{0.94\columnwidth}
        \footnotesize
        \ResultPass: successful goal arrival;
        \ResultFail{}: unsuccessful run.
        Failure codes: C, collision; D, deadlock; T, timeout.
        Environments are numbered from left to right and
        from top to bottom in Fig.~\ref{fig:s1_run}.
    \end{minipage}
\end{table}

Fig.~\ref{fig:s1_run} presents twelve selected S1 environments, and Table~\ref{tab:static_outcomes} reports the corresponding navigation outcomes. The proposed method completes all selected tasks, including routes through narrow passages and successive turns.

Fig.~\ref{fig:s1_process} shows the corridor updates in a representative run. Local SCPs are constructed along the upcoming reference path and combined into a safety corridor. The corridor changes as the robot moves, providing guidance through successive turns.

Across all 25 S1 environments, the proposed method achieves a $100\%$ success rate and a mean navigation time of $30.98\,\mathrm{s}$. The complete comparison and representative baseline failures are provided in the appendix.

\subsubsection{Dynamic Environments}

S2 evaluates navigation when moving obstacles change the available free space.
Fig.~\ref{fig:s2_process} shows a representative sequence. The corridor is rebuilt using updated obstacle positions and the locally feasible reference segment. The robot adjusts its motion to the changing obstacle configuration and continues toward the goal.

The proposed method completes all 25 S2 tasks without collision, with a mean navigation time of $19.80\,\mathrm{s}$. Table~\ref{tab:complete_navigation_results} summarizes the success rates, failure categories, and mean navigation times for both environment sets. These results demonstrate performance in the tested environments, rather than safety under arbitrary obstacle motions.

\begin{figure}[!tbp]
    \centering
    \includegraphics[width=\columnwidth]
    {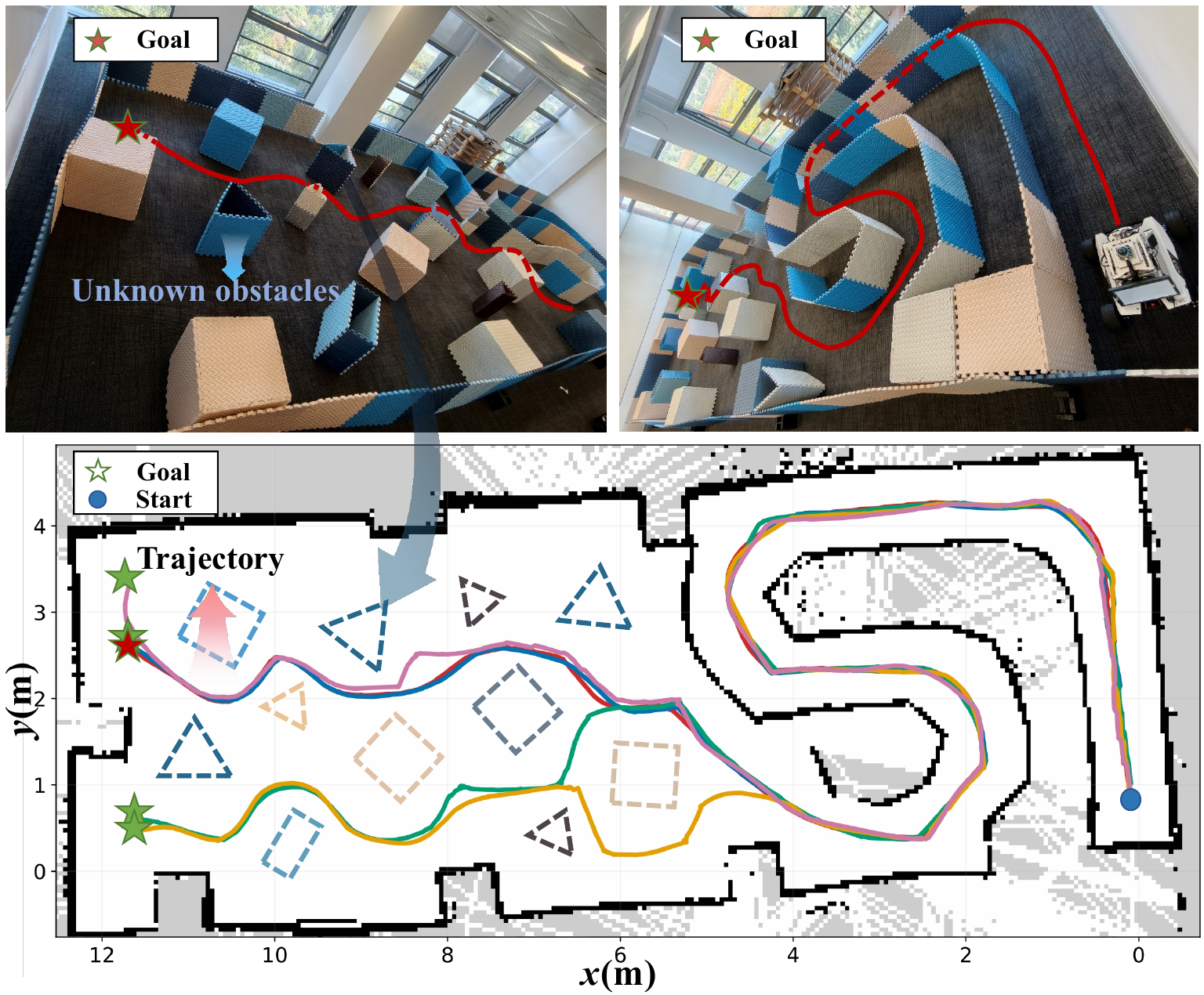}
    \caption{
    The experimental environments and the executed trajectories from five navigation runs.
    The upper-left and upper-right pictures show the unknown obstacle region and the narrow passage, matching the left and right parts of the map below. The red curves in the photos illustrate the paths related to the red goal markers on the map. In the lower panel, solid curves in different colors show the five recorded trajectories. The blue circles are the starting points, and stars mark the goals. Dashed outlines indicate obstacles absent from the prior map.}
    \label{fig:vehicle_trajectories}
\end{figure}

\begin{figure*}[!t]
    \centering
    \includegraphics[width=\textwidth]
    {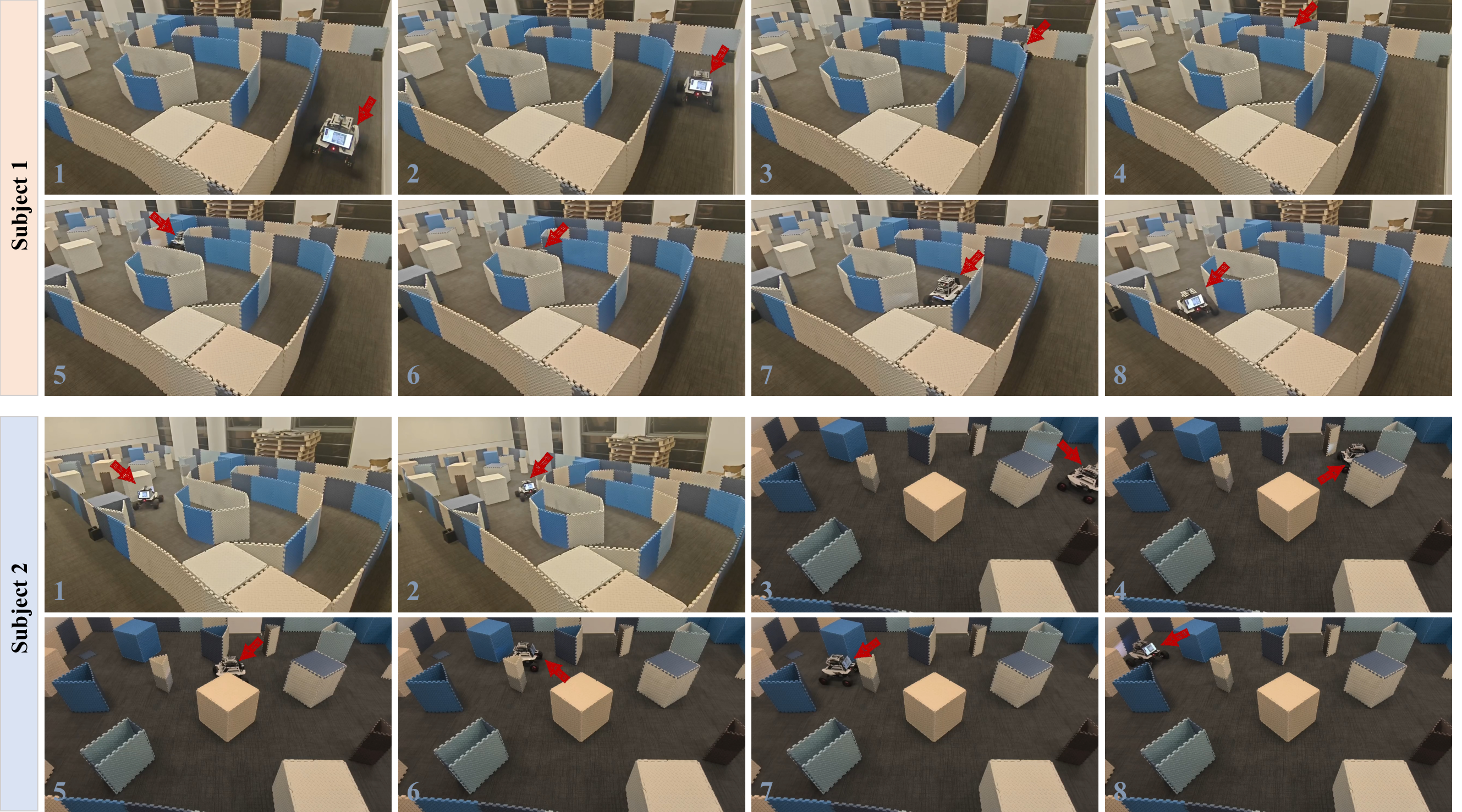}
    \caption{
    Real-world navigation through narrow passages and unknown obstacles. 
    The upper half shows Subject 1, in which the vehicle goes through a narrow passage with continuous turns. The lower half shows Subject 2, where the vehicle enters and passes through a region with unknown obstacles. Snapshots within each subject are arranged chronologically from left to right, top to bottom. The red arrows indicate the vehicle's position.}
    \label{fig:vehicle}
\end{figure*}
\subsection{Ground Vehicle Real-World Demonstration (Experiments)}
\label{subsec:experiments}
We evaluate the proposed framework on the four-wheel differential-drive vehicle shown in the lower-right
panel of Fig.~\ref{fig:framework}. The vehicle is equipped with a Livox MID-360 LiDAR and measures $580~\mathrm{mm} \text{ W} \times 612~\mathrm{mm} \text{ L}$. The experiments are conducted in a $13~\mathrm{m} \times 5~\mathrm{m}$ indoor test area. 

The planar motion of the vehicle is approximated by
\begin{equation}
    \dot{x} = v\cos\theta,\qquad
    \dot{y} = v\sin\theta,\qquad
    \dot{\theta} = \omega,
    \label{eq:ground_vehicle_model}
\end{equation}
where $(x,y)$ is the vehicle position, $\theta$ is its heading, and $v$ and $\omega$ are the linear and angular velocity inputs. Online LiDAR observations are used to update the reference path and the local safety corridor.

To account for vehicle size and orientation, the four corners and four edge midpoints of a rectangular footprint enlarged by a safety clearance are evaluated in the Poisson field. The field values and directional derivatives at these points provide feedback for simultaneous speed reduction and steering correction. The feedback is smoothed, and the resulting commands are limited to $0\le v\le0.50\,\mathrm{m/s}$ and $|\omega|\le0.60\,\mathrm{rad/s}$. The control loop runs at a nominal rate of $50\,\mathrm{Hz}$. Two subjects are considered: navigation through a narrow passage and navigation among obstacles absent from the prior map.

\subsubsection{Subject 1: Narrow-Passage Navigation}
As shown in the upper half of Fig.~\ref{fig:vehicle},
Subject 1 consists of narrow passages, with continuous curves and changes in direction. The limited lateral clearance requires the vehicle to coordinate forward motion and turning.

The look-ahead corridor shows the available space along the path the vehicle is about to travel. Its Poisson field helps with local reference generation and footprint-based feedback. The image sequence shows the vehicle moving through the passage and making continuous turns. The trajectories on the right side of Fig.~\ref{fig:vehicle_trajectories} show the complete routes the vehicle took through this region across the recorded runs.

\subsubsection{Subject 2: Navigation Among Unknown Obstacles}
As shown in the lower half of Fig.~\ref{fig:vehicle}, Subject 2 contains obstacles of different sizes and orientations beyond the passage exit. All obstacles in this region are absent from the prior map. The vehicle can only observe their geometry through online LiDAR observations. Therefore, the path planned based on the prior map might collide with obstacles in the actual environment.

When an obstacle is detected, the observations are used to update the path and the local safety corridor. The image sequence shows the vehicle entering the area and navigating between obstacles. The left side of Fig.~\ref{fig:vehicle_trajectories} shows how the obstacles are laid out and the routes to different goals. 

Together, the two subjects examine navigation under geometric constraints and incomplete prior obstacle information. The five recorded trajectories in Fig.~\ref{fig:vehicle_trajectories} show how the vehicle moves through narrow passages and navigates among previously unknown obstacles to reach different goals. These results demonstrate the applicability of the proposed framework to a real ground vehicle with nonholonomic motion constraints and a rectangular chassis.

\section{Conclusions}
This paper presented a planning-guided construction of look-ahead safety corridors and their interior fields for BaS-based navigation. Adaptive compensation reduces shrinkage relative to fixed compensation while retaining containment within the union of local safe regions. Poisson reconstruction preserves the prescribed corridor boundary and provides an interior field for one geometric barrier state. The analysis establishes corridor invariance under barrier-state boundedness and active-corridor validity.

The proposed method completed all tasks in 25 static and 25 dynamic simulation environments. Experiments with a differential-drive ground vehicle further demonstrate navigation through narrow passages and among previously unknown obstacles. Future work will investigate safety guarantees under time-varying obstacle motion and tighter integration between path planning and corridor construction.

\appendix
\subsection{Barrier-State Dynamics}
\label{app:bas_dynamics}
The Barrier-State dynamics follow the construction
in~\cite{almubarak2025barrier}. Let $\beta(x)=B(h(x))$.
Strict monotonicity of $B$ ensures that $B^{-1}$ exists
on $B((0,\infty))$. The composition $B'\circ B^{-1}$ is assumed to be locally Lipschitz on this range.

For $z+\beta_e\in B((0,\infty))$, the functions in the augmented system~\eqref{eq:augmented_system} are
\begin{align}
    f_b(x,z)
    ={}&
    \bigl(B'\circ B^{-1}\bigr)(z+\beta_e)L_fh(x)
    \nonumber\\
    &-\gamma\bigl(z+\beta_e-\beta(x)\bigr),
    \label{eq:app_bas_fb}\\
    g_b(x,z)
    ={}&
    \bigl(B'\circ B^{-1}\bigr)(z+\beta_e)L_gh(x),
    \label{eq:app_bas_gb}
\end{align}
where $\gamma>0$ is an observer-like gain. The correction term acts on the discrepancy between $z+\beta_e$ and $\beta(x)$.

According to the BaS consistency result in~\cite[Lemma~1]{almubarak2025barrier}, the initialization
\begin{equation}
    z(0)=B(h(x(0)))-\beta_e
    \label{eq:app_bas_initialization}
\end{equation}
preserves~\eqref{eq:bas_consistency} along the well-posed
evolution in the safe interior. On this consistent evolution, the correction term vanishes and the Barrier-State dynamics reduce to
\begin{equation}
    \dot z
    =B'(h(x))
    \bigl(L_fh(x)+L_gh(x)u\bigr).
    \label{eq:app_bas_consistent_dynamics}
\end{equation}
This is the time derivative of $B(h(x))-\beta_e$.

The consistency relation also applies to an admissible
equilibrium $x_e$ other than the origin by choosing $\beta_e=B(h(x_e))$. Admissibility means that the system can remain at $x_e$ under an admissible steady input. The conditional safety result only requires $\beta_e$ to be finite and fixed during each active interval.

\subsection{Complete Simulation Results and Failure Cases}
\label{app:simulation_results}

Table~\ref{tab:complete_navigation_results} summarizes all 200 runs.
In S1, JPS and JPS-BaS both achieve a $40\%$ success rate,
but their failure modes differ.
JPS has 15 collisions, whereas JPS-BaS has 12 deadlocks
and three timeouts.
JPS-SCP-BaS achieves $84\%$, with four timeouts.
The proposed method completes all 25 tasks.

In S2, the success rates are $16\%$, $92\%$, $92\%$,
and $100\%$ for JPS, JPS-BaS, JPS-SCP-BaS,
and the proposed method, respectively.
JPS has 21 collisions.
JPS-BaS has one deadlock and one timeout.
JPS-SCP-BaS has one collision and one timeout.

The proposed method has the lowest mean navigation
time among the four methods in both sets.
However, these means use each method's successful
runs, which may correspond to different environments.
They should therefore be interpreted together
with the success rates, rather than as a paired
comparison of navigation speed.

\begin{figure}[!tbp]
    \centering
    \includegraphics[width=\columnwidth]
    {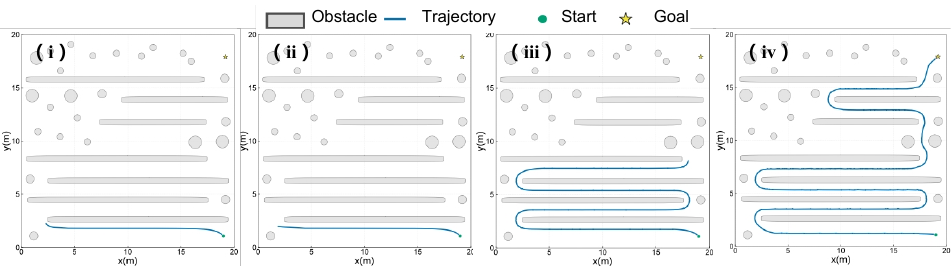}
    \caption{
    Representative navigation outcomes in S1: 
    (i) JPS, collision (C); (ii) JPS-BaS, deadlock (D); (iii) JPS-SCP-BaS, timeout (T); and (iv) the proposed method, successful. All panels use the same environment, start, and goal. Blue curves show the executed robot-center trajectories up to termination. Gray regions denote obstacles, green circles indicate the start, and yellow stars mark the goal.}
    \label{fig:baseline_failures}
\end{figure}

Fig.~\ref{fig:baseline_failures} illustrates the three
failure categories in a representative S1 environment.
JPS terminates in collision near the first turn.
JPS-BaS triggers the deadlock criterion before passing
that turn.
JPS-SCP-BaS passes several turns but does not reach
the goal within the time limit.
The proposed method completes the route.
These examples illustrate distinct failure modes;
the aggregate results are given in
Table~\ref{tab:complete_navigation_results}.

\begin{table}[!tbp]
    \centering
    \caption{Navigation outcomes and mean navigation time.}
    \label{tab:complete_navigation_results}
    \renewcommand{\arraystretch}{1.12}
    \setlength{\tabcolsep}{3pt}
    \footnotesize
    \begin{tabular*}{\columnwidth}
        {@{\extracolsep{\fill}}lccccc@{}}
        \hline
        Method & Success & C & D & T
        & \shortstack{Time\\(s)}\\
        \hline
        \multicolumn{6}{c}{S1: Static environments}\\
        \hline
        JPS
            & 10/25 (40\%) & 15 & 0 & 0 & 31.23\\
        JPS-BaS
            & 10/25 (40\%) & 0 & 12 & 3 & 31.99\\
        JPS-SCP-BaS
            & 21/25 (84\%) & 0 & 0 & 4 & 38.54\\
        Ours
            & 25/25 (100\%) & 0 & 0 & 0 & 30.98\\
        \hline
        \multicolumn{6}{c}{S2: Dynamic environments}\\
        \hline
        JPS
            & 4/25 (16\%) & 21 & 0 & 0 & 21.03\\
        JPS-BaS
            & 23/25 (92\%) & 0 & 1 & 1 & 25.32\\
        JPS-SCP-BaS
            & 23/25 (92\%) & 1 & 0 & 1 & 24.57\\
        Ours
            & 25/25 (100\%) & 0 & 0 & 0 & 19.80\\
        \hline
    \end{tabular*}

    \vspace{3pt}
    \begin{minipage}{\columnwidth}
        \footnotesize
        Success: successful runs/total runs (success rate).
        C: collision; D: deadlock; T: timeout.
        Time: mean simulated start-to-goal time over
        successful runs only, excluding computation time.
        Ours: JPS-Corridor-BaS.
    \end{minipage}
\end{table}

\bibliographystyle{IEEEtran}
\bibliography{bibligraphy}
\end{document}